\documentclass[12pt]{article}
\usepackage{fancyhdr,graphicx,amsmath,amssymb, amsthm}
\usepackage{graphicx,psfrag,epsf}
\usepackage{enumerate}
\usepackage{comment}
\usepackage{url}
\usepackage{amsfonts} 
\usepackage{hyperref} 
\usepackage{bm}

\usepackage{subcaption}
\usepackage[dvipsnames]{xcolor}
\usepackage{float} 
\usepackage{enumitem}
\usepackage{authblk}
\usepackage{pifont}
\usepackage{wrapfig}

\usepackage[top=1.25in, bottom=0.75in, left=1in, right=1in]{geometry}

\usepackage{sectsty}
\newcommand{\spacingset}[1]{%
  \renewcommand{\baselinestretch}{#1}\small\normalsize}

\usepackage[style=apa]{biblatex}
\usepackage{titlesec}
\titleformat{\section}{\normalfont\normalsize\bfseries}{\thesection.}{1em}{\centering}
\titleformat{\subsection}
  {\normalfont\normalsize}{\thesubsection.}{1em}{}
\titleformat{\subsubsection}[runin]
  {\normalfont\normalsize\itshape}{\thesubsubsection.}{1em}{}

\newtheorem{theorem}{Theorem}[section]
\newtheorem{corollary}[theorem]{Corollary}

\newtheorem{lemma}[theorem]{Lemma}

\newtheorem{remark}[theorem]{Remark}
\usepackage{algpseudocode}
\usepackage{algorithm}
\usepackage{pgfplots}
\pgfplotsset{compat=1.18}
\usepgfplotslibrary{groupplots}
\algrenewcommand\algorithmicrequire{\textbf{Input:}}
\algrenewcommand\algorithmicensure{\textbf{Output:}}

\pgfplotsset{
  every axis plot/.append style={line width=0.5pt, no markers},
  grid style/.append style={line width=0.2pt},
  every axis/.append style={line width=0.4pt} 
}

\pgfplotsset{
  every axis/.style={
    grid=none,                 
    tick align=outside,
    label style={font=\small},
    title style={font=\small},
    legend style={draw=none, fill=none, font=\small},
    scale only axis,
    minor tick num=0,
    scaled x ticks=false,
    ticklabel style={/pgf/number format/fixed}
  },
  tight ticks/.style={
    tick align=outside,
    xticklabel style={yshift=-1.5pt},   
    yticklabel style={xshift=-1.5pt},   
    ticklabel style={inner sep=0pt},  
    major tick length=2pt,            
    minor tick length=1pt,             
    ticklabel style={font=\scriptsize}
  },
  fxplot1/.style={width=2.5cm, height=2.5cm, xmin=-0.05, xmax=0.05, ymin=0, ymax=2, xtick={-0.05,0,0.05}, ytick={0,1,2}},
  fxplot2/.style={width=2.5cm, height=2.5cm, xtick={-0.04,-0.02,0,0.02,0.04}, xmin=-0.05, xmax=0.05, ymin=-2, ymax=0, xtick={-0.05,0,0.05}, ytick={-2,-1,0}},
  fxplot3/.style={width=6cm, height=4cm, xtick={-0.04,-0.02,0,0.02,0.04}, xmin=-0.045, xmax=0.045, ymin=0.5, ymax=1.5, ytick={0.5,1,1.5}},
  fxplot4/.style={width=2.5cm, height=2.5cm, xtick={-0.4,-0.2,0,0.2,0.4},ytick={0,0.5,1,1.5}, xmin=-0.4, xmax=0.4, ymin=0, ymax=1.5},
  fxplot5/.style={width=2.5cm, height=2.5cm, xtick={-0.4,-0.2,0,0.2,0.4},ytick={-1.5,-1,-0.5,0}, xmin=-0.4, xmax=0.4, ymin=-1.5, ymax=0},
  fxplot6/.style={width=6cm, height=4cm, xtick={-1000,0,500},ytick={0.6,0.8,1}, xmin=-1300, xmax=700, ymin=0.5, ymax=1.1},
  psiplot/.style={width=6cm, height=4cm, xmin=0, xmax=1, ymin=-1, ymax=1,xtick={0,1},ytick={-1,1}},
  mapplot/.style={width=2.5cm, height=2.5cm, xmin=0, xmax=1, ymin=0, ymax=1,xtick={0,1},ytick={0,1}},
  phiplot/.style={width=2.5cm, height=2.5cm, xmin=0, xmax=1, ymin=-0.12, ymax=0.12, ytick={-0.1,0,0.1}},
  densplot1/.style={width=2.5cm, height=2.5cm, xmin=0, xmax=1, ymin=0, ymax=8},
  densplot2/.style={width=6cm, height=4cm, xmin=0, xmax=1, ymin=0, ymax=3.2,xtick={0,0.2,0.4,0.6,0.8,1}},
  pplot/.style={width=6cm, height=4cm, xmin=0, xmax=1, ymin=0, ymax=1,xtick={0,0.2,0.4,0.6,0.8,1},ytick={0,1}},
  dispplot/.style={width=6cm, height=4cm, xmin=0, xmax=1, ymin=0, ymax=0.15,xtick={0,0.2,0.4,0.6,0.8,1},ytick={0,0.1}},
  picplot/.style={width=2.5cm, height=2.5cm, xmin=0, xmax=1, ymin=0, ymax=1,xtick={0,1},ytick={0,1}}
}

\newcommand{\rd}{\mathrm{d}}

\DeclareMathOperator{\id}{id}
\DeclareMathOperator{\E}{E}

\title{Kantorovich Regression Analysis of Random Distributions with Mixed Predictors}

\author[1]{Kaheon Kim}
\author[1]{Changbo Zhu\thanks{Zhu's research was supported by NSF DMS-2412832.}}
\affil[1]{University of Notre Dame, Department of ACMS}
\date{}

\begin{document}
\spacingset{1}

\maketitle

\begin{abstract}
In modern scientific applications, observations often take the form of probability distributions. We study regression problems with distribution-valued responses and mixed distributional and Euclidean predictors. Under quadratic optimal transport, the proposed Kantorovich regression model approximates the response Kantorovich potential from the Wasserstein barycenter as a linear combination of transformed predictor Kantorovich potentials, where the transformations are determined by one-dimensional functional parameters. The formulation naturally accommodates mixed predictors, allowing Euclidean covariates to enter as scaling coefficients, and extends to multivariate distributions. We characterize functional parameter classes ensuring the intrinsic structure of the model, and  establish asymptotic consistency of model parameters through the convex Sobolev loss. Real data applications include a mixed-predictor analysis of housing price distributions and an analysis of two-dimensional temperature distributions, demonstrating the flexibility and interpretability of the proposed framework.
\end{abstract}


\newpage
\spacingset{1.8}
\section{Introduction}
Distributions arise naturally in many scientific disciplines as fundamental objects for summarizing information. Rather than reducing observations to a few summary statistics such as the mean or selected quantiles, probability distributions provide a comprehensive description of variability, asymmetry, and multimodality. When each observational unit is associated with a probability distribution, this naturally leads to the analysis of distribution-valued data, where each distribution is treated as a single data point [\cite{petersen2019wasserstein,zhang2011functional,panaretos2020invitation,pana:16, PETERSEN2022159}]. Specific examples include age-at-death distributions at the country level [\cite{Jdanov2021, shang2017grouped}], yearly temperature distributions [\cite{bhat:21}],  individual-level distributional profiles of physical activity [\cite{matabuena2023distributional}], and subject-specific density estimates for intra-hub brain connectivity [\cite{petersen2016functional}], among others. 

Optimal transport has emerged as a geometrically interpretable framework for analyzing distribution-valued data. Its geometric nature stems from the fact that the optimal transport map moves mass from one distribution to another in the most efficient way, minimizing the total cost of transportation among all mass-preserving maps. Under quadratic cost, the optimal transport map $T$ from a source distribution to a target distribution induces a vector field defined by $ x \mapsto T(x) - x$, which we refer to as the displacement field. The Wasserstein barycenter, as a notion of average for distributions [\cite{AguehCarlier2011}], satisfies that the average displacement fields from the barycenter to the sample distributions vanish, paralleling the role of the mean in linear spaces. Unlike pointwise averaging of densities, which can produce artificial multimodality when distributions differ mainly by shifts, the Wasserstein barycenter preserves the overall geometric structure of the distributions by balancing the transport displacements. 

In this paper, we study the regression problem where the response is distribution-valued, and the predictors can be distributions, vectors, or both. By leveraging the displacement field encoded by Kantorovich potentials relative to Wasserstein barycenters, we develop a regression framework, termed Kantorovich regression, that models how predictor-induced displacement fields combine to explain variation in the response displacement field. To incorporate distributional predictors, the key mechanism is to scale each predictor displacement field through a one-dimensional functional parameter over the level sets of the corresponding potential function. Equivalently, a one-dimensional functional parameter is composed with the Kantorovich potential. This mechanism allows more flexibility than existing approaches that use scalar coefficients and focus primarily on one-dimensional distributions [\cite{zhu2023atm, chen2024distribution, zhu2025geodesic}].  We then aggregate the scaled displacement fields and subtract an intercept term so that the resulting displacement field is centered, allowing it to approximate the response displacement field relative to the barycenter. 

Our design of the scaling operation is motivated by two considerations. First, it makes the model intrinsic over the chosen parameter space, as the predicted transport maps remain cyclically monotone without requiring projection. Lemma~\ref{lem:connectedcomponents} shows that preserving cyclic monotonicity requires the functional parameter to be constant on each connected component of the level sets of the potential function. Since identifying these connected components is generally difficult, particularly in multivariate settings, we instead impose uniform scaling over each level set. Methods that utilize the manifold structure of Wasserstein space [\cite{chen2023wasserstein, zhan:20}], or more generally map distributions to a Hilbert space [\cite{kokoszka2019forecasting, petersen2016functional}], are considered extrinsic. Second, it provides computational advantages, as the scaling parameter is always one-dimensional regardless of the dimensionality of the distributions. Following the computational framework in \cite{jacobs2021back, kim2025optimal, kim2025sobolev}, we estimate the model parameters by minimizing the Sobolev loss using first-order gradient descent algorithms, where the distributions and parameter functions are discretized over a regular grid for numerical computation.

For Euclidean predictors, we model each predictor as a uniform scaling factor applied to a parameter displacement field, defined as the gradient of a $c$-concave potential function. The scaled displacement fields from distributional predictors and the parameter displacement fields scaled by Euclidean predictors are added together. Another line of methods, originally proposed by \cite{ghod:21} and later extended by \cite{chen2026multi, ghodrati2024distributional}, focuses on one-dimensional distributions by learning parametric transport maps. It remains unclear how to extend this framework to the mixed predictor setting. To provide a clearer overview of the methodological differences, we summarize and compare the key features of selected representative methods in Table \ref{tab:method_comparison}.
\begin{table}[ht]
    \centering
    \begin{tabular}{l|ccccccc}
        & WR & DoD & GOT   & DIDO & MT & GNW & KR \\ \hline
        Functional parameter 
        & \ding{51} & \ding{51} & \ding{55} & \ding{55} & \ding{51} & N/A & \ding{51} \\
        Intrinsic 
        & \ding{55} & \ding{51} & \ding{51} & \ding{55}& \ding{51} & \ding{51} &  \ding{51} \\
        Multivariate distributions & \ding{55} & \ding{55} & \ding{51} & \ding{55} & \ding{55} & \ding{51} & \ding{51}  \\
        Multiple distributional predictors 
        & \ding{51} & \ding{55} & \ding{51} & \ding{51}& \ding{51} & \ding{51} & \ding{51} \\
        Mixed predictors 
        & \ding{55} & \ding{55} & \ding{55} & \ding{55} & \ding{55} & \ding{55} & \ding{51}
    \end{tabular}
    \caption{Comparison of key features across methods, including whether functional or scalar regression coefficients are used, intrinsic preservation of OT structure, applicability to multivariate distributions, accommodation of multiple distributional predictors, and support for mixed distributional and scalar predictors. WR: \cite{chen2023wasserstein}; DoD: \cite{ghod:21}; GOT: \cite{zhu2025geodesic}; DIDO: \cite{chen2024distribution}; MT: \cite{chen2026multi}; GNW: generalized Nadaraya-Watson [\cite{gnw}];  KR: the proposed Kantorovich regression model.}
    \label{tab:method_comparison}
\end{table}

\section{Preliminary}\label{sec:pre}
Throughout the manuscript, let $\Omega \subset \mathbb{R}^d$ be a compact set. For $x,y \in \Omega$, let $\langle x,y \rangle = x^\top y$ denote the Euclidean inner product, and $\|x\| = \sqrt{\langle x,x\rangle}$ the induced Euclidean norm on $\mathbb{R}^d$. The diameter of $\Omega$ is defined as $\mathrm{diam}(\Omega) := \sup_{x, y
 \in \Omega} \| x- y \| < \infty$. We write $\mathcal{C}^{k}(\Omega)$ for the space of
real-valued functions $\phi:\Omega\to\mathbb{R}$ whose partial derivatives up to
order $k$ exist and are continuous on $\Omega$, and use $\nabla \phi (x), \nabla^2 \phi (x)$ to represent its gradient vector and Hessian matrix at $x$ respectively. The homogeneous Sobolev norm w.r.t a probability measure $\nu$ on $\Omega$ is defined as $\| \phi \|_{\dot{H}^1(\nu)} : = ( \int \| \nabla \phi (x) \|^2  \rd \nu(x))^{1/2}$. For symmetric matrices $A,B\in\mathbb{R}^{d\times d}$, $A \preceq B $ means that $B-A$ is positive semidefinite (PSD). 

We let $\mathcal{P}_{\text{ac}}(\Omega)$ be the set of absolutely continuous probability measures on $\Omega$. For $\mu \in \mathcal{P}_{\text{ac}}(\Omega)$, its support is denoted as $\mathrm{sp}(\mu)$. For any cost function $c:\Omega \times \Omega \rightarrow \mathbb{R}_{+}$, the $c$-transform of a function $\phi:\Omega\to\mathbb{R}$ is defined as 
$
\phi^c(y) = \inf_{x\in\Omega}\{ c(x, y) -\phi(x)\}.
$
In this manuscript, we focus on the quadratic cost $c(x, y) =  \| x-y \|^2/2$. Letting $T_\#\mu$ denote the pushforward measure of $\mu$ by $T:\Omega\to\Omega$, the Monge transportation problem is formulated as 
\begin{align} 
\inf_{T:\,T_\#\mu=\nu}\, \int_\Omega c(T(x), x) \, \rd \mu(x), \qquad \text{(MP)}\label{eq:W2-MP} 
\end{align} 
where the minimization is over the set of maps $T:\Omega \rightarrow \Omega$ such that $T_{\#}\mu = \nu$. If $\mu, \nu \in \mathcal{P}_{\mathrm{ac}}(\Omega)$, the solution $T$ to the Monge 
problem is $\mu$-a.e.\ unique. This map is called the \emph{optimal transport map} 
from $\mu$ to $\nu$, and is denoted by $T_{\mu\to\nu}$. In this case, the 
quadratic Wasserstein distance between $\mu$ and $\nu$ is $W_2(\mu,\nu)
:=  \left( 2 \int_\Omega c(x, T_{\mu\to\nu}(x) )\, \mathrm{d}\mu(x) \right)^{1/2}$. For discrete measures, the Monge problem often has no solution because it does not allow mass to be split in order to match the target weights. In contrast, the Kantorovich (relaxed) formulation always admits a solution, 
and defines
$$
\inf_{\pi \in \Pi(\mu,\nu)}
\int_{\Omega \times \Omega}
\frac{1}{2}\|x - y\|^2 \, \mathrm{d}\pi(x,y),
$$
where $\Pi(\mu,\nu)$ denotes the set of all couplings (joint distributions) on $\Omega \times \Omega$ with marginals $\mu$ and $\nu$. For $A,B\subseteq\mathbb R^d$ and a function $\psi:B\to\mathbb R$,
define $\psi^c(x)=\inf_{y\in B}\{c(x,y)-\psi(y)\}, x\in A$. A function $\phi:A\to\mathbb R$ is called $c$-concave relative to
$(A,B)$ if $\phi=\psi^c$ for some $\psi:B\to\mathbb R$. We denote
the collection of such functions by $c\text{-}\mathrm{conc}(A,B)$
and write $c\text{-}\mathrm{conc}(\Omega)
:=c\text{-}\mathrm{conc}(\Omega,\Omega)$. For $\mu,\nu\in\mathcal P_{\mathrm{ac}}(\Omega)$, let
$A=\operatorname{sp}(\mu)$ and $B=\operatorname{sp}(\nu)$. The dual
Kantorovich formulation can then be written as
\begin{align}
\sup_{\phi\in c\text{-}\mathrm{conc}(A,B)}
\left\{
\int_A \phi\,\rd\mu
+
\int_B \phi^c\,\rd\nu
\right\}.
\qquad \text{\emph{(DP)}}
\label{eq:W2-DP}
\end{align}

For absolutely continuous measures, we have 
$
\inf \text{ (MP)} = \sup \text{ (DP)}. 
$
Any maximizer of (DP) is called a \emph{Kantorovich potential}, and is $c$-concave and unique $\mu$-a.e. up to an additive constant. In statistical analysis, a unique representative, denoted by $\phi_{\mu\to\nu}$, can be selected by requiring it to satisfy a normalization condition, such as $
\int_{\Omega} \phi_{\mu\to\nu}(x)\, \rd \mu(x)=0$. The optimal transport map and the Kantorovich potential satisfy $T_{\mu\to\nu}(x) = x - \nabla \phi_{\mu\to\nu}(x),  x \in \Omega$, and this motivates us to define
$$
T_{\phi} := \id - \nabla \phi \text{ for any } c\text{-concave } \phi.
$$
Moreover, when $\phi$ is $c$-concave, the function $\tfrac{\|x\|^2}{2} - \phi(x)$ is convex (see \cite{santambrogio2015optimal, villani2009optimal}), which implies that $T_{\mu\to\nu} = \nabla u$ for some convex function $u:\Omega\to\mathbb{R}$. This result is commonly referred as the Brenier theorem (\cite{Brenier_polarization_1991}). A map $T : \mathbb{R}^d \to \mathbb{R}^d$ is said to be cyclically monotone if, for any integer $m \ge 2$ and any collection of points $\{x_1,\dots,x_m\} \subset \mathbb{R}^d$, 
$
\sum_{i=1}^m \langle T(x_i),\, x_i - x_{i+1} \rangle \ge 0, x_{m+1} := x_1.
$
A classical result of Rockafellar (\cite{rockafellar1966characterization}) states 
that $T$ is cyclically monotone if and only if there exists a proper, lower 
semicontinuous convex function 
$\varphi : \mathbb{R}^d \to \mathbb{R} \cup \{+\infty\}$ such that
$
T(x) \in \partial \varphi(x) \text{ for all } x \in \mathbb{R}^d,
$
where $\partial \varphi$ denotes the subdifferential of $\varphi$. This implies that $T_{\mu \rightarrow \nu}$ between absolutely continuous $\mu$ and $\nu$ is cyclically monotone. 

Let $P_{\mu}$ be a probability measure on $\mathcal{P}_{\mathrm{ac}}(\Omega)$, 
and let $\mu \sim P_{\mu}$ denote a random measure drawn from it. 
The (population) Wasserstein barycenter of $P_{\mu}$, denoted $\bar{\mu}$, is 
defined as the unique minimizer of the barycenter functional $B:\mathcal{P}_{\mathrm{ac}} (\Omega) \rightarrow \mathbb{R}$, where $
  B (\omega) :=   \int_{\mathcal{P}_{\mathrm{ac}}(\Omega)} W_2^2(\omega, \mu')\, \rd P_{\mu} (\mu').$ Given i.i.d.\ samples $\mu_1,\dots,\mu_n \sim P_{\mu}$, the empirical Wasserstein 
barycenter $\bar{\mu}_n$ is the unique minimizer of $
\widehat{B}(\omega) := \frac{1}{n} \sum_{i=1}^n W_2^2(\omega, \mu_i)$. First-order gradient-based methods have recently been developed by \cite{kim2025optimal} and \cite{kim2025sobolev} for accurate computation of Wasserstein barycenters, leveraging Wasserstein and Sobolev gradients to yield efficient algorithms with convergence guarantees.

\section{Kantorovich Regression} \label{sec:KR}
\subsection{Composition operation on displacement field}
Given two distributions $\mu_1, \mu_2 \in \mathcal{P}_{\mathrm{ac}}  (\Omega)$, we recall that the gradient of the Kantorovich potential $\nabla \phi_{\mu_1\to\mu_2}$ and the optimal transport $T_{\mu_1\to\mu_2}$ are connected through
\[
-\nabla \phi_{\mu_1\to\mu_2}(x) \;=\; T_{\mu_1\to\mu_2}(x) - x,
\]
so $-\nabla\phi_{\mu_1\to\mu_2}(x)$ is the displacement vector from each location $x$ in the source distribution to its optimally matched location in the target distribution. In addition, if $\bar\mu$ denotes the Wasserstein barycenter of a random distribution $\mu$, then the average displacement at each $x$ satisfies
\[
\E_\mu\!\left[\nabla \phi_{\bar\mu\to\mu}(x)\right] = 0.
\]
The barycenter serves as the mean of $\mu$, and the displacement field $\nabla\phi_{\bar\mu\to\mu}$ encodes how $\mu$ departs from $\bar\mu$, which directly parallels the Euclidean case $\E [X - \E[X]] = 0$ for any random vector $X \in \mathbb{R}^d$. In the ordinary regression model $ Y - \E[Y] = A ( X - \E[X] ) + \epsilon $, multiplying by the model parameters $A \in \mathbb{R}^{d\times d}$ can be interpreted as an adjustment step for the displacement vector $  X - \E[X] $. If $A$ is a scalar, then the adjustment is uniform across all entries of $X - \E[X]$. Since $-\nabla \phi_{\bar\mu \to \mu}(x)$ and $\nabla \phi_{\bar\mu \to \mu}(x)$ differ only by a sign, which does not affect the modeling framework, we henceforth focus on $\nabla \phi_{\bar\mu \to \mu}(x)$ for convenience.

Motivated by the analogies, we construct our regression model by first proposing a non-uniform adjustment of the displacement field $\nabla\phi_{\bar\mu\to\mu}$. In particular, given a function $f \in \mathcal{C}^2(\mathbb{R})$ and a $c$-concave potential $\phi:\Omega\to\mathbb{R}$, we define
\begin{equation}\label{eq:circleop}
    f \circledcirc \phi := \begin{cases}
        
    f \circ \phi, &\text{ if } f \text{ is non-decreasing and concave}; \\
    -f \circ (-\phi), &\text{ if } f \text{ is non-increasing and convex}.
    \end{cases}
\end{equation}
The function $f$ can be seen as the functional parameter, and by applying this operation on $\phi_{\bar\mu \rightarrow \mu}$, the adjusted displacement at $x$ becomes 
\[
\nabla(f\circledcirc \phi_{\bar\mu \rightarrow \mu})(x) = \begin{cases}
    f'(\phi_{\bar\mu \rightarrow \mu}(x))\,\nabla\phi_{\bar\mu \rightarrow \mu}(x)&\text{ if } f \text{ is non-decreasing and concave}. \\
    f'(-\phi_{\bar\mu \rightarrow \mu}(x))\nabla\phi_{\bar\mu \rightarrow \mu}(x)&\text{ if } f \text{ is non-increasing and convex}.
\end{cases} 
\]

\begin{figure}[t]
\centering
\includegraphics[width=0.8\textwidth]{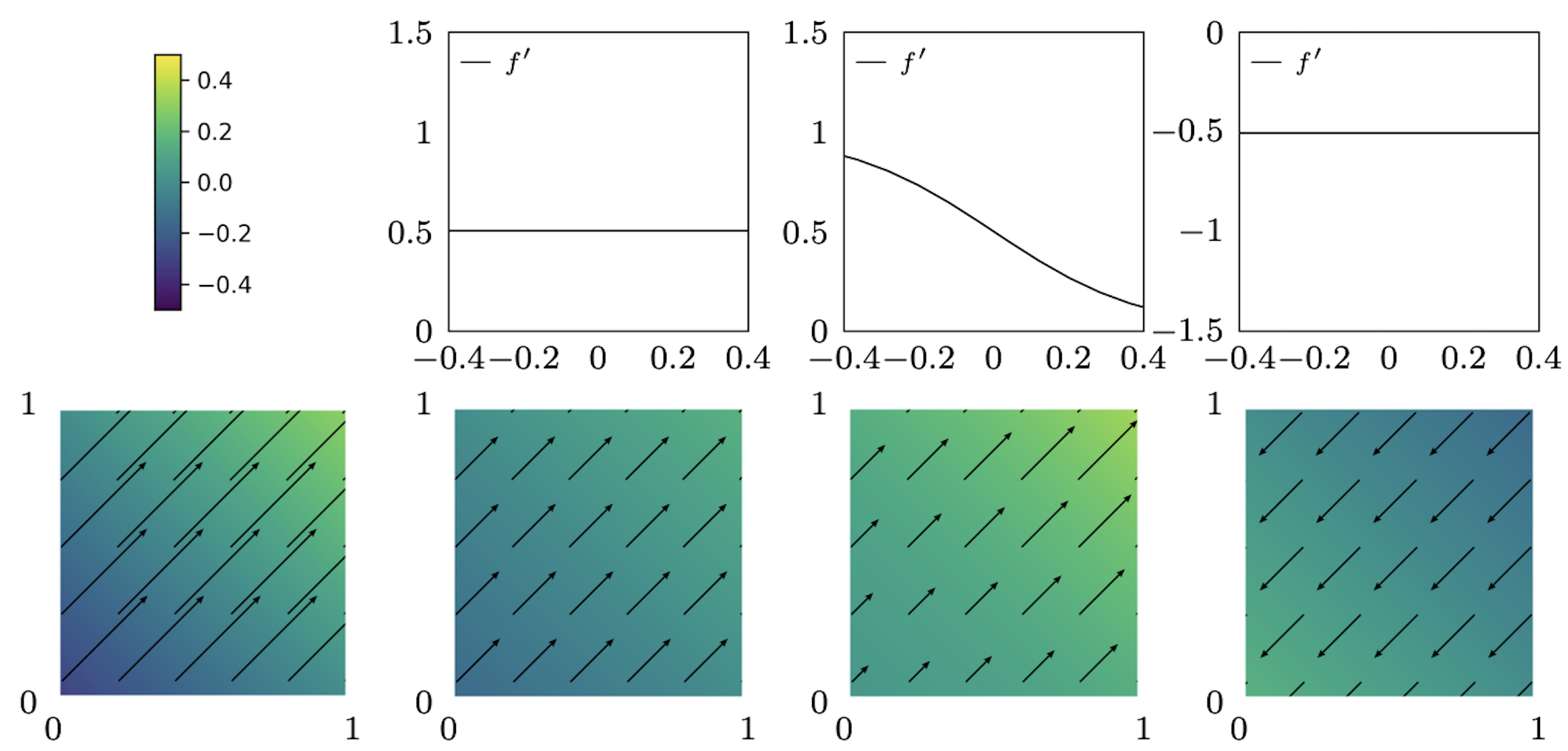}
\caption{Illustration of the adjustment on $\phi$ by $f$. Column 1 (bottom row) shows the Kantorovich potential $\phi:[0, 1]^2 \rightarrow \mathbb{R}$ as a heat map, together with the gradient field $\nabla \phi$ drawn as arrows. From column~2 onward, the top row displays the derivatives $f'$ of the functional parameter $f$, while the bottom row shows $f\circledcirc\phi$ as heat maps with the corresponding gradient field $\nabla (f\circledcirc\phi)$ (arrows).}
\label{fig:multiplication}
\end{figure}

When $f$ is non-decreasing and concave, the displacement direction remains $\nabla\phi_{\bar\mu \rightarrow \mu}(x)$, while the length is rescaled by $f'(\phi_{\bar\mu \rightarrow \mu}(x))\geq0$.
Points sharing the same level set $\{x:\phi_{\bar\mu \rightarrow \mu}(x)=a\}$ receive identical scaling $f'(a)$, and the scaling varies 
across level sets.
When $f$ is non-increasing and convex, $- f'(-\phi_{\bar\mu \rightarrow \mu}(x))\geq 0$.
The direction becomes opposite to $\nabla\phi_{\bar\mu \rightarrow \mu}(x)$ and its magnitude is rescaled by $- f'(-\phi_{\bar\mu \rightarrow \mu}(x))$.
Again, points sharing a level set $\{x:-\phi_{\bar\mu \rightarrow \mu}(x)=b\}$ receive the same non-negative multiplier $-f'(b)$ and the scaling varies 
from one level set to the next. When $f(t) = a t$ for some fixed $a \in \mathbb{R}$, $f \circledcirc \phi = a \phi$ regardless of the sign of $a$ and this adjustment of displacement vector $\nabla \phi (x)$ by $f$ is uniform in $x$. Figure \ref{fig:multiplication} illustrates how the function $f$ adjusts a two-dimensional potential $\phi$ and its corresponding displacement field.

We recall that valid optimal transport maps are cyclically monotone. The lemma below states that suppose one attempts to modify the displacement field by replacing 
$\nabla\phi(x)$ with $g(x)\,\nabla\phi(x)$ for some function 
$g:\Omega\to\mathbb{R}$. In order for the resulting 
map $x \mapsto x - g(x)\,\nabla\phi(x)$ to remain cyclically monotone, 
the function $g$ must be constant on each connected component of every 
level set of $\phi$. Since identifying the connected components of the level sets of a potential function is difficult, adjustments of the form $T_{f \circ \phi}(x):= x - f' (\phi(x)) \nabla \phi(x)$ offer a natural and convenient construction of optimal transport.

\begin{lemma}\label{lem:connectedcomponents}
Given a $c$-concave function $\phi \in \mathcal{C}^2(\Omega)$ such that $\nabla \phi(x) \neq 0$ for any $x \in \Omega$ and $g\in \mathcal{C}^1(\Omega)$, we let $R(x) := x - g (x) \nabla \phi (x) $. If $R$ is cyclically monotone, then $g$ is constant on each connected component of each level set $S_a = \{ x : \phi(x) = a\}$.
\end{lemma}

\subsection{Single distributional predictor}\label{subsec:single_dist}
We first consider regression of the distributional response on a single
distributional predictor. Let $P_{\nu, \mu}$ be a joint distribution on $\mathcal{P}_{\text{ac}}(\Omega) \times \mathcal{P}_{\text{ac}}(\Omega)$, $(\nu, \mu) \sim P_{\nu, \mu}$, and $\{ (\nu_i, \mu_i) \}_{i=1}^n$ be i.i.d.\ copies of $(\nu, \mu)$. The marginal distributions on $\nu$ and $\mu$ are denoted as $P_{\nu}$ and $P_{\mu}$ respectively, and their population Wasserstein barycenters are $\bar\nu$ and $\bar\mu$. We use $\E_{\mu_i}[\cdot]$ to denote expectation with respect to the randomness of $\mu_i$. For instance, for any functional $F:\mathcal{P}_{\mathrm{ac}}(\Omega)\to\mathbb{R}$, $
\E_{\mu_i}[F(\mu_i)]
=
\int F(\omega)\, \rd P_{\mu}(\omega).$

In practice, distributions are estimated from samples, and their supports are difficult to determine. A convenient and statistically effective approach is to estimate all distributions on a common domain using standard procedures such as kernel density estimation. Typical choices
of the common domain include a bounded
interval $[a,b]$ for one-dimensional distributions, or a rectangle
$[a,b]\times[c,d]$ for two-dimensional distributions. In such cases,
all distributions involved in the regression analysis share the same
support. Motivated by this practice, we consider throughout the paper the case
in which the barycenters of the predictor and response distributions
have identical support, i.e.,  $\mathrm{sp}(\bar\nu) = \mathrm{sp} (\bar\mu)$. This setting includes the common support scenario described above as a
special case. The handling of the situation in which
$\mathrm{sp}(\bar\nu) \neq \mathrm{sp} (\bar\mu)$ is discussed separately
in Remark \ref{rmk:sp}.

Because the supports of the observed distributions and the gradients
of the predicted potentials are uniformly bounded under enforced regularity conditions, we take $\Omega$
to be a sufficiently large compact convex set containing the supports
of all observed and predicted distributions. For a predicted potential $\phi$, let
$
A=\operatorname{sp}(\bar\nu) \text{ and }
B=T_\phi(A).
$
Since $\phi$ matters only on $A$, we call $\phi$ $c$-concave if its
restriction to $A$ is $c$-concave relative to $(A,B)$. By construction,
$A,B\subseteq\Omega$.

The key idea of our Kantorovich regression is to explain variability in the response displacement field 
$\nabla \phi_{\bar{\nu} \rightarrow \nu_i}$ through adjustments of the predictor displacement field 
$\nabla \phi_{\bar{\mu} \rightarrow \mu_i}$ via the operation defined in Equation~\ref{eq:circleop}, and models the conditional mean of response Kantorovich potential as
\begin{align}\label{eq:single}
\E[ \phi_{\bar\nu \rightarrow \nu_i}(x) \, | \, \mu_i ] =    f  \circledcirc\phi_{\bar\mu\to\mu_i}(x) -\E_{\mu_i} \left[f \circledcirc \phi_{\bar\mu\to\mu_i} (x) \right],
\end{align} 
where $f:\mathbb{R} \rightarrow \mathbb{R}$ is the  functional parameter that is either non-decreasing and concave or non-increasing and convex. Here, $f \circledcirc \phi_{\bar\mu\to\mu_i}$ denotes the adjusted potential function, which is further centered by subtracting the intercept term $ \E_{\mu_i} [f \circledcirc \phi_{\bar\mu\to\mu_i} (x) ]  $. When $f(t) = a t$ for some $a \in \mathbb{R}$, the operation $f \circledcirc \phi_{\bar\mu\to\mu_i}$ is linear in $\phi_{\bar\mu\to\mu_i}$, and the intercept term vanishes, i.e.,
$\E_{\mu_i} [f \circledcirc \phi_{\bar\mu\to\mu_i} (x) ] = 0$.
For a general (nonlinear) functional parameter $f$, this centering is necessary to ensure that
$
\E [ T_{\,f \circledcirc \phi_{\bar\mu\to\mu_i} - \E_{\mu_i} [f \circledcirc \phi_{\bar\mu\to\mu_i}]}(x) ] = x$. For one-dimensional distributions, the induced optimal transport map is monotone and thus allows for a more structured modeling of the error [\cite{ghod:21, zhu2023atm}]. In this case, we can write the model as
\begin{align} \label{kr:1d}
T_{\bar\nu \to \nu_i} = T_{\varepsilon_i} \circ T_{\,f \circledcirc \phi_{\bar\mu\to\mu_i} - \E_{\mu_i} [f \circledcirc \phi_{\bar\mu\to\mu_i}]},
\end{align}
where $T_{\varepsilon_i}:\Omega \rightarrow \mathbb{R}$ is a random increasing map such that $\E[T_{\varepsilon_i} (x)] = x$.

For a matrix $A \in \mathbb{R}^{m \times m}$, the operator norm induced by the Euclidean norm is defined as
$
\|A\|_{\mathrm{op}} := \sup_{\|x\| = 1} \|Ax\|$. We center the potential function $\phi_{\bar\mu \rightarrow \mu_i}$ so that $ \int \phi_{\bar\mu \rightarrow \mu_i} \rd \bar\mu = 0$ and consider the set
\begin{align}
\mathbb{K} (\gamma_-, \gamma_+) := \Big\{\phi\in  \mathcal{C}^2(\Omega) :  \| \nabla \phi (x) \| \leq \mathrm{diam}(\Omega),  -\gamma_- I_d \preceq \nabla^2 \phi(x) \preceq \gamma_+ I_d  \Big\}.
\end{align}
The set $\mathbb{K} (\gamma_-, \gamma_+ )$ contains potential functions whose curvatures are uniformly controlled by $\gamma_-$ and $\gamma_+$. We note that any twice differentiable $c$-concave function $\varphi$ is $1$-smooth, i.e., $ \nabla^2 \varphi (x)\preceq I_d$ and thus $\gamma_+ \leq 1$. In contrast, the constant $\gamma_-$ may be large. We choose the interval $[a,b]$ to be sufficiently large so that it contains the ranges of both $\phi_{\bar\mu\to\mu_i}(x)$ and $-\phi_{\bar\mu\to\mu_i}(x)$ for all $i$. Let 
$$
\mathbb{F}_{+}(\kappa_1,\kappa_2) := \{f\in \mathcal{C}^2 ([a, b]) : 0 \leq f'(t) \le \kappa_1,\ -\kappa_2 \leq  f''(t) \le 0, \ \forall \, t \in [a, b] \}
$$ 
be a class of non-decreasing concave functional parameters and  
$
\mathbb{F}_{-}(\kappa_1,\kappa_2) := \{f\in \mathcal{C}^2 ([a, b]) : 0 \leq -f'(t) \le \kappa_1,\ -\kappa_2 \leq  -f''(t) \le 0, \ \forall \, t \in [a, b] \} $ be the set of non-increasing convex functions. The constants $\kappa_1, \kappa_2$ can be selected, depending on the data-generating distribution $P_{\mu,\nu}$, so that the centered potential
$f \circledcirc \phi_{\bar\mu\to\mu_i} - \E \left[f \circledcirc \phi_{\bar\mu\to\mu_i}\right]$
is $c$-concave. Consequently, the associated transport map $T_{\,f \circledcirc \phi_{\bar\mu\to\mu_i} - \E [f \circledcirc \phi_{\bar\mu\to\mu_i}]}$ is cyclically monotone. 

\begin{theorem}\label{thm:single_concavity}
Let $\eta$ and $\lambda$ be population-level constants defined as
\begin{align*}
\eta := \sup_{x\in \mathrm{sp}(\bar\nu)} \E_{\mu_i}[\|\nabla\phi_{\bar\mu\to\mu_i}(x)\|^2] \text{ and }   \lambda := \sup_{x\in \mathrm{sp}(\bar\nu)} (\E_{\mu_i}[\| \nabla^2 \phi_{\bar\mu\to\mu_i} (x) \|_{\mathrm{op}}^2 ])^{1/2}.
\end{align*}
Assuming that $\phi_{\bar\mu\to\mu_i} \in \mathbb{K} (\gamma_-, \gamma_+)$ almost surely, then for each $\delta \in \{+, -\}$, by choosing $\kappa_{1, \delta}, \kappa_{2, \delta}$ to satisfy 
\begin{equation}\label{single_condition}
  (\gamma_{\delta} + \lambda ) \kappa_{1,\delta}+ \eta \kappa_{2, \delta} \leq 1, 
\end{equation}
the function $f\circledcirc\phi_{\bar\mu\to\mu_i}-\E_{\mu_i} \left[f\circledcirc \phi_{\bar\mu\to\mu_i}\right]$ is $c$-concave for any $f \in \bigcup_{\delta \in \{+, -\} } \mathbb{F}_{\delta}(\kappa_{1, \delta}, \kappa_{2, \delta}) $.
\end{theorem}
If the sign indicator $\delta=+$, we fit a model with a non-decreasing concave functional parameter. If $\delta=-$, we fit a model with a non-increasing convex functional parameter. The constants $\eta, \lambda, \gamma_{-}, \gamma_{+}$ depend on the generating distribution $P_{\nu,\mu}$ and can be estimated from data. In particular, $\eta$ represents the maximum displacement on average. Although $\lambda \leq \max \{ \gamma_-, \gamma_+ \}$, the value of $\lambda$ can be substantially smaller, since it is a population-level quantity defined through an expectation. If the optimal transport corresponds to a uniform displacement, i.e., $T_{\phi_{\bar\mu\to\mu_i}}(x) = x + a_i$ for some $a_i \in \mathbb{R}^d$, then $\nabla^2 \phi_{\bar\mu\to\mu_i} (x) = 0$. Consequently, we may take $\gamma_- = \gamma_+ = \lambda = 0$, and the constants $\kappa_{1, -}, \kappa_{1, +}$ can be chosen arbitrarily large. Condition \eqref{single_condition}  can always be made true by selecting $\kappa_{1, +}, \kappa_{2, +}, \kappa_{1, -}, \kappa_{2, -}$ small enough. 

When $f(t) = a t$ for some constant $a \in \mathbb{R}$, we have $f \circledcirc \phi_{\bar\mu\to\mu_i} = a \phi_{\bar\mu\to\mu_i}$, the intercept term $\E[f \circledcirc \phi_{\bar\mu\to\mu_i}(x)] = 0$ and model~\eqref{eq:single} reduces to
\begin{align*}
\E[ \phi_{\bar\nu \rightarrow \nu_i}(x) \, | \, \mu_i ] =  a \phi_{\bar\mu\to\mu_i} (x).
\end{align*}
In this case, we can set $\kappa_{2, +} = \kappa_{2, -} = 0$, and the sufficient conditions in the preceding theorem simplify accordingly, yielding the following corollary.

\begin{corollary}\label{cor:single_concavity}
Let $f(t) = a t$ for some $a \in \mathbb{R}$. 
If $\phi_{\bar\mu\to\mu_i} \in \mathbb{K} (\gamma_-, \gamma_+)$ almost surely, then by choosing $\kappa_{1, +}$ and $\kappa_{1, -}$ to satisfy $ \kappa_{1, +} \gamma_{+} \leq 1 $ and $ \kappa_{1, -} \gamma_{-} \leq 1 $ respectively, the function $f \circledcirc \phi_{\bar\mu\to\mu_i} = a \phi_{\bar\mu\to\mu_i}$ is $c$-concave for any $ - \kappa_{1, -} \le a \le \kappa_{1, +} $.
\end{corollary} 

The functional parameter $f$ is infinite dimensional and, for computational purposes, must be discretized on a fixed grid
$a = z_1 < z_2 < \cdots < z_K = b$
that covers the union of the images of $\{\phi_{\bar\mu\to\mu_i}\}$.
When $f$ is non-decreasing and concave, its derivative $f'$ is nonnegative and non-increasing.
We can approximate $f'$ using a stepwise function of the form
$\sum_{k=1}^K \theta_k \mathbf{1}_{\{x \le z_k\}}$,
where $\theta_k \ge 0$ are parameters, and the discretization error can be controlled by refining the grid, that is, by increasing $K$. Furthermore, we let
\begin{equation}\label{parametric_func}
f'_\theta(x)
=
\sum_{k=1}^K \frac{\theta_k}{1 + \exp\bigl(\theta_0 (x - z_k)\bigr)}
\end{equation}
be a smooth approximation to the stepwise function $\sum_{k=1}^K \theta_k \mathbf{1}_{\{x \le z_k\}}$. It can be seen that as $\theta_0 \to \infty$, $f'_\theta(x)$ converges to
$\sum_{k=1}^K \theta_k \mathbf{1}_{\{x \le z_k\}}$.

\begin{figure}[t]
\centering
\includegraphics[scale=0.25]{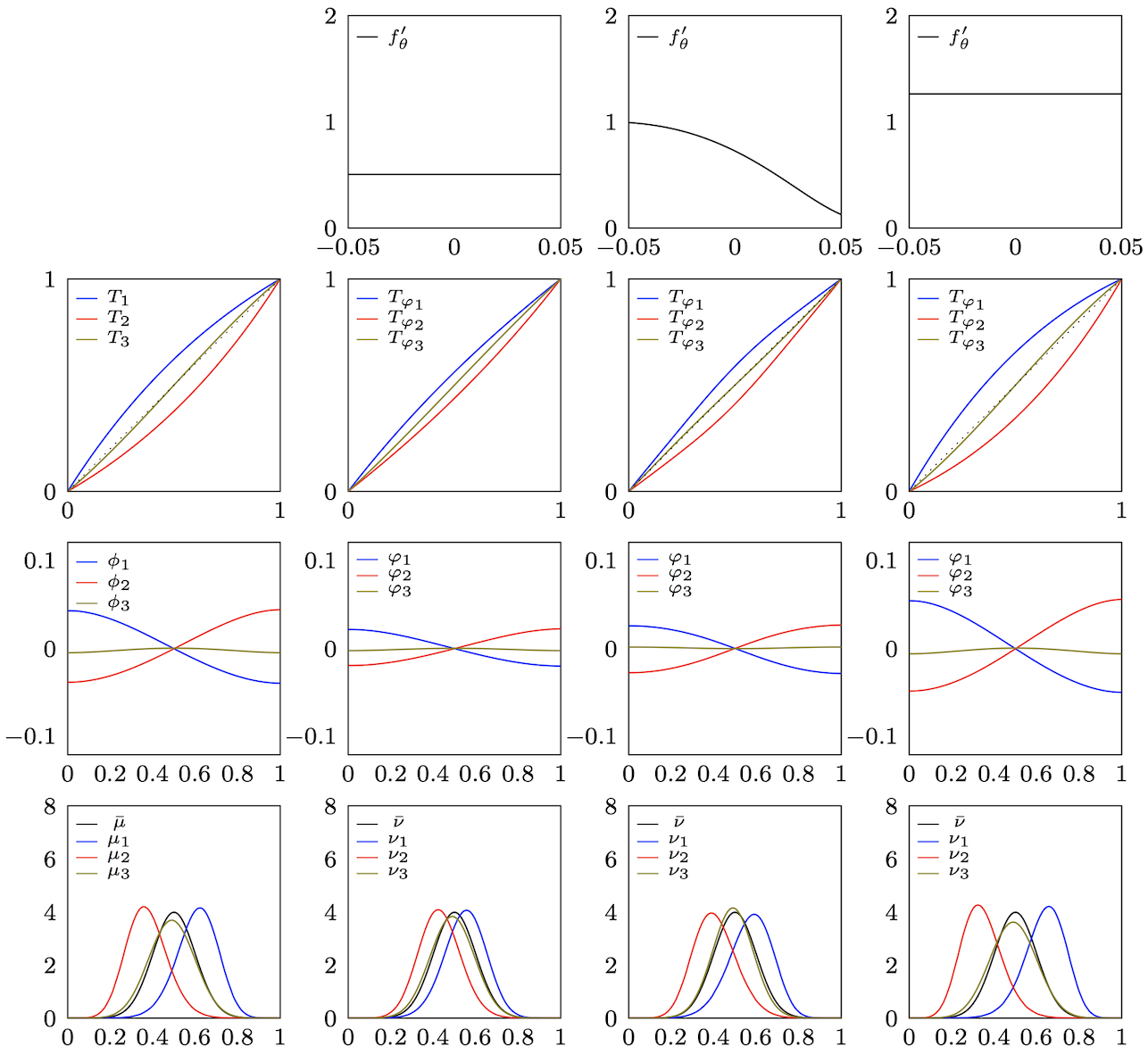}
\caption{1D illustration of KR model with non-decreasing concave functional parameter.}
\label{fig_demo1D+}
\end{figure}

We next provide one-dimensional and two-dimensional examples with  functional parameter $f_{\theta}$ that satisfy the assumptions in Theorem \ref{thm:single_concavity}. For a 1D example, we fix $\bar\mu$ as normal distribution $N(0.5, 0.1^2)$ truncated on the interval $[0,1]$, and construct $\mu_i = (T_i)_{\#} \bar\mu$ for $i=1,2,3$, where 
$
T_1(x)=\frac{1-e^{-x}}{1-e^{-1}}, \,
T_2(x)=\frac{e^{x}-1}{e-1} \text{ and }T_3(x)=3x-T_1(x)-T_2(x).
$
We set \(-0.05=z_1<\cdots<z_{100}=0.05\) that cover the range of the \(\phi_{\bar\mu \rightarrow \mu_i}\) and consider three parameter settings of $\{\theta_k\}$: (1) \(\theta_k = 2\times10^{-4}\,k\), \(\theta_0=0\) so that \(f_\theta(t)=0.505 \, t \); (2) \(\theta_k = 2\times10^{-4}\,k\), \(\theta_0=100\), and it holds that \(f_\theta\in \mathbb{F}_+(0.9925, 13.39) \); (3) \(\theta_k = 5\times10^{-4}\,k\), \(\theta_0=0\) and \(f_\theta(t) = 1.2625 \, t \). We then set $\bar\nu = \bar\mu$, $\varphi_i = f_{\theta} \circ \phi_{\bar\mu \rightarrow \mu_i} - \sum_{j=1}^3 f_{\theta} \circ \phi_{\bar\mu \rightarrow \mu_j}/3 $ and $\nu_i = (T_{\varphi_i})_{\#} \bar\nu$. These distributions and potential functions are plotted in Figure \ref{fig_demo1D+}. It can be computed that $\eta = (\sqrt{e}-1)^{2} / (6(\sqrt{e}+1)^{2}) \approx 9.998\times10^{-3}$, $\lambda = \sqrt{2e^2-10e+14}/(\sqrt{3}(e-1)) \approx0.4244$, and $\phi_{\bar\mu\to\mu_i}\in \mathbb{K} (\gamma_-, \gamma_+)$ with $\gamma_- =  1 / (e-1 ) \approx 0.5820$, $\gamma_+ = (e-2)/(e-1) \approx 0.4180$, thus $f_\theta$ under settings (1) and (3) satisfy the condition in Corollary \ref{cor:single_concavity}, and under setting (2) satisfies the condition (\ref{single_condition}). 

From Figure \ref{fig_demo1D+}, we observe when $f$ is given as a non-decreasing and concave function, the relative locations of $\mu_1,\mu_2,\mu_3$ with respect to $\bar\mu$ are preserved in $\nu_1,\nu_2,\nu_3$ with respect to $\bar\nu$: $\mu_1$ and $\nu_1$ are on the right, $\mu_2$ and $\nu_2$ are on the left, and $\mu_3$ and $\nu_3$ are in the center. Under setting (1) (second column of Figure \ref{fig_demo1D+}), $f_{\theta}(t) = 0.505t$, the distributions $\nu_1$ and $\nu_2$ move closer to $\bar\nu$ than $\mu_1$ and $\mu_2$ are to $\bar\mu$. Conversely, in the fourth column $f_{\theta}(t) = 1.2625t$, $\nu_1$ and $\nu_2$ are farther from $\bar\nu$ than $\mu_1$ and $\mu_2$ are from $\bar\mu$. The third column  for setting (2) exhibits the same locational effects as (1). However, $\nu_3$ lies above $\bar\nu$ while $\mu_3$ lies below $\bar\mu$, a discrepancy caused by the local deformation induced by the nonlinear $f_{\theta}$, which cannot occur when the functional parameter $f$ is linear. The experiment repeated with the functional parameter set to $-f_{\theta}$ is provided in the supplementary material. 

\begin{figure}[t]
\centering
\includegraphics[scale=0.2]{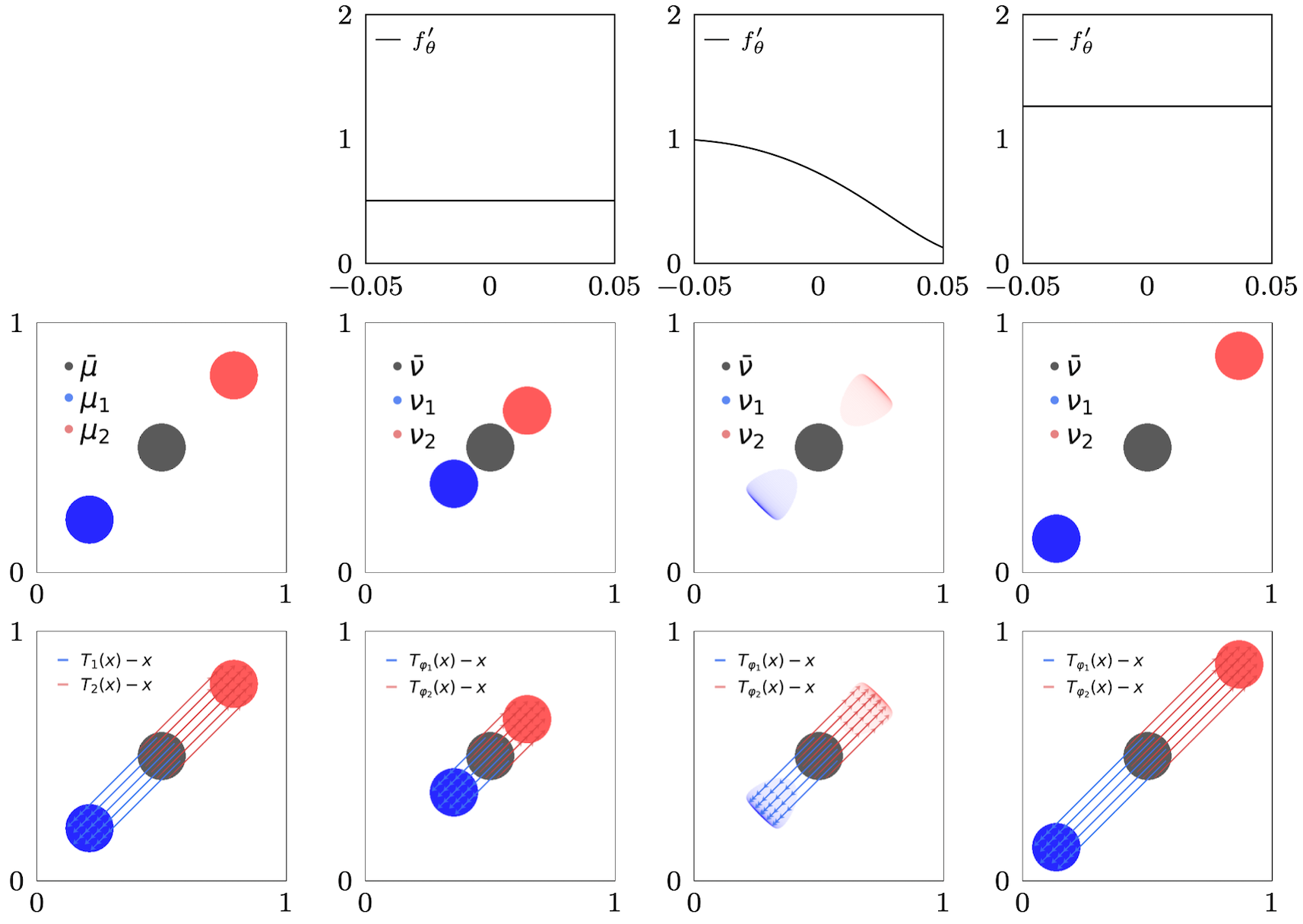}
\caption{2D illustration of KR with non-decreasing concave functional parameter.}
\label{fig:2D+}
\end{figure}

For a 2D example, we let $\mu_1, \mu_2$ be uniform distributions supported on round disks of diameter 0.1 centered at $(0.2, 0.2)$ and $(0.8, 0.8)$ respectively. The Wasserstein barycenter $\bar\mu$ of $\mu_1, \mu_2$ is uniform on disk centered at $(0.5, 0.5)$, and we set \(\bar\nu = \bar\mu\). To illustrate the Kantorovich regression model, we use the same set of functions $f_{\theta}$ and compute $\varphi_i = f_{\theta} \circledcirc \phi_{\bar\mu \rightarrow \mu_i} - \frac{1}{2} \sum_{j=1}^2 f_{\theta} \circledcirc \phi_{\bar\mu \rightarrow \mu_j} $ and  $\nu_i = ( T_{\varphi_i})_{\#} \bar\nu$. In Figure \ref{fig:2D+}, where \(f'\) is non-decreasing and concave, the relative positions
of \(\mu_1,\mu_2\) with respect to \(\bar\mu\) are preserved by \(\nu_1,\nu_2\)
with respect to \(\bar\nu\): \(\mu_1\) and \(\nu_1\) lie in the lower left, and
\(\mu_2\) and \(\nu_2\) lie in the upper right. In the second column, with
\(f_\theta(t)=0.505 t\), \(\nu_1\) and \(\nu_2\) are
closer to \(\bar\nu\) than \(\mu_1\) and \(\mu_2\) are to \(\bar\mu\). In the fourth column, with \(f_\theta(t)= 1.2625 t \) , \(\nu_1\) and
\(\nu_2\) are farther from \(\bar\nu\) than \(\mu_1\) and \(\mu_2\) are from
\(\bar\mu\). In these cases, the function \(f\) scales the length of the
displacement vector \(T_i(x)-x\) uniformly, yielding uniform distributions. By
contrast, in third column, scale $f'_{\theta}(\phi_{\bar\mu\to\mu_i} (x))$ varies across $x \in \mathrm{sp} (\bar\nu )$, so the resulting distributions are not uniform. 

\begin{remark} \label{rmk:sp}
We now discuss the case when $\mathrm{sp}(\bar\nu) \neq \mathrm{sp}(\bar\mu)$. By choosing a fixed translation $\tau: \mathrm{sp}(\bar\nu) \rightarrow \mathrm{sp}(\bar\mu)$, model~\ref{eq:single} and Theorem~\ref{thm:single_concavity} can be applied in the same way by replacing $\phi_{\bar\mu\to\mu_i}$ with $\phi_{\bar\mu \rightarrow \mu_i} \circ \tau$. The only difference is that the constants appearing in Theorem~\ref{thm:single_concavity} must now account for the effect of $\tau$. For one-dimensional distributions, if all distributions are supported on connected intervals, then the barycenter is also supported on an interval. Let $\mathrm{sp}(\bar\nu) = [a,b]$ and $\mathrm{sp}(\bar\mu) = [c,d]$. We may choose $\tau$ to be the affine transformation mapping $[a,b]$ to $[c,d]$. In this case, composing with $\tau$ does not change the shape of $\phi_{\bar\mu \rightarrow \mu_i}$, so the Kantorovich model continues to deform $\phi_{\bar\mu \rightarrow \mu_i}$ only through the model parameters.
\end{remark}

\begin{remark} \label{rmk:f}
In model~\ref{eq:single}, we impose structural constraints on $f$, requiring it to be either non-decreasing and concave or non-increasing and convex. These constraints are motivated by the conditions in Theorem~\ref{thm:single_concavity}. In principle, one could allow any $f \in \mathcal{C}^2(\Omega)$ as the functional model parameter and consider the model
\begin{align*}
\E[ \phi_{\bar\nu \rightarrow \nu_i}(x) \, | \, \mu_i ] =  f \circ \phi_{\bar\mu\to\mu_i}(x) - \E_{\mu_i}[f \circ \phi_{\bar\mu\to\mu_i} (x) ].
\end{align*} 
The trade-off is that ensuring $ f \circ \phi_{\bar\mu\to\mu_i} - \E_{\mu_i}[f \circ \phi_{\bar\mu\to\mu_i}] $ remains $c$-concave would require stronger bounds on both $f'$ and $f''$, together with uniform bounds on $\|\nabla\phi_{\bar\mu\to\mu_i}(x)\|$ and $\|\nabla^2\phi_{\bar\mu\to\mu_i}(x)\|_{\mathrm{op}}$ over all $i$ and $x$, compared to those used in Theorem~\ref{thm:single_concavity}.
\end{remark}

\subsection{Multiple distributional predictors}

We now extend the Kantorovich regression (KR) model to incorporate multiple
distributional predictors. Let $(\nu, \bm{\mu}) \sim P_{\nu,\bm{\mu}}$, where
$\bm{\mu} = (\mu^{1}, \ldots, \mu^{p}) \in (\mathcal{P}_{\mathrm{ac}}(\Omega))^{p}$,
and $P_{\nu,\bm{\mu}}$ is a probability measure on
$(\mathcal{P}_{\mathrm{ac}}(\Omega))^{p+1}$. Given an i.i.d.\ sample $\{(\nu_i, \bm{\mu}_i)\}_{i=1}^n$ drawn from
$P_{\nu,\bm{\mu}}$, we write $\bm{\mu}_i = (\mu_i^{1}, \ldots, \mu_i^{p})$.
Let $P_\nu$ and $P_{\mu^j}$ denote the marginal distributions of $\nu$
and $\mu^j$, respectively, and let $\bar{\nu}$ and $\bar{\mu}^j$ denote
their corresponding population Wasserstein barycenters. Similar to the one predictor case, we consider $\mathrm{sp}(\bar\nu) = \mathrm{sp}(\bar\mu^j)$ for all $j$, and define the Kantorovich regression (KR) model with multiple distributional predictors as follows:
\begin{equation}\label{eq:multi}
\E[ \phi_{\bar\nu \rightarrow \nu_i}(x) \, | \, \bm{\mu}_i ] = \sum_{j=1}^p \varphi_i^j(x),
\end{equation}
where $f_j$ is the functional parameter and  $\varphi_i^j (x)  =  f_j \circledcirc \phi_{\bar\mu^j\to\mu_{i}^j} (x) - \E_{\mu^j_i} [f_j \circledcirc \phi_{\bar\mu^j\to\mu_{i}^j}(x)] $. Likewise, for one-dimensional distributions, we can model the random error by composing with a random increasing function $T_{\varepsilon_i}$ with $\E[T_{\varepsilon_i}(x)] = x$ and consider the model 
\begin{align}\label{kr:mult1d}
T_{\bar\nu \rightarrow \nu_i} =  T_{\varepsilon_i} \circ
T_{ \sum_{j=1}^p \varphi_i^j}.
\end{align}
There are $p$ functional parameters, each of which is either non-decreasing and concave or non-increasing and convex. We encode this structure by $\bm{\delta} = (\delta_1,\dots,\delta_p)^\top \in \{+,-\}^p$, where $\delta_j = +$ indicates that the $j$th parameter is non-decreasing and concave, and $\delta_j = -$ indicates that it is non-increasing and convex. Let 
$$
\eta^j := \sup_{x\in \mathrm{sp}(\bar\nu)} \E_{\mu_i^j}[\|\nabla\phi_{\bar\mu^j\to\mu_{i}^j}(x)\|^2], \quad \lambda^j := \sup_{x\in \mathrm{sp}(\bar\nu)} (\E_{\mu_i^j}[\| \nabla^2 \phi_{\bar\mu^j\to\mu_{i}^j} (x) \|_{\mathrm{op}}^2 ])^{1/2}
$$ 
and for simplicity, we set $\bm{\lambda} = (\lambda^1, \lambda^2, \dots, \lambda^p)^\top$, $\bm{\eta} = (\eta^1, \eta^2, \dots, \eta^p)^\top$, $ \bm{\kappa}_{1,\bm{\delta}}  = (\kappa_{1, \bm{\delta} }^1, \dots, \kappa_{1, \bm{\delta}}^p)^\top$, $ \bm{\kappa}_{2,\bm{\delta}}  = (\kappa_{2, \bm{\delta} }^1, \dots, \kappa_{2, \bm{\delta} }^p)^\top$, $\bm{\gamma}_{\bm{\delta}} = (\gamma^1_{\delta_1}, \dots, \gamma^p_{\delta_p})^\top \in \mathbb{R}^p$ and
\begin{align*}
    \bm{\mathbb{F}}_{\bm{\delta}}(\bm{\kappa}_{1, \bm{\delta}}, \bm{\kappa}_{2, \bm{\delta}}) = \mathbb{F}_{\delta_1} (\kappa_{1, \bm{\delta}}^1, \kappa_{2, \bm{\delta}}^1  ) \times \dots \times \mathbb{F}_{\delta_p} (\kappa_{1, \bm{\delta}}^p, \kappa_{2, \bm{\delta}}^p ).
\end{align*}
Here, each component $\kappa_{1, \bm{\delta}}^j$  and   $\kappa_{2, \bm{\delta}}^j$ depends on the entire vector $\bm{\delta}$. Similarly, we center the potential function $\phi_{\bar\mu^j \rightarrow \mu_i^j}$ so that $ \int \phi_{\bar\mu^j \rightarrow \mu_i^j} \rd \bar\mu^j = 0 $ and choose the interval $[a,b]$ sufficiently large to contain the ranges of $\phi_{\bar\mu^j\to\mu_{i}^j}$ and $-\phi_{\bar\mu^j\to\mu_{i}^j}$ for all $i, j$. 

\begin{corollary}\label{cor:multiple_concavity}
Assuming $\phi_{\bar\mu^j\to\mu_{i}^j}\in\mathbb{K} (\gamma^{j}_-, \gamma^{j}_+)$ almost surely, then for each $\bm{\delta} \in \{ + , -\}^p $, by choosing $ \bm{\kappa}_{1, \bm{\delta}}, \bm{\kappa}_{2, \bm{\delta}} $ to satisfy
\begin{equation*}
(\bm{\gamma}_{\bm{\delta}}^\top + \bm{\lambda}^\top ) \bm{\kappa}_{1, \bm{\delta}}  + \bm{\eta}^\top \bm{\kappa}_{2, \bm{\delta}}  \leq 1,
\end{equation*}
the function $\sum_{j=1}^p \varphi_i^j$ is $c$-concave for any $( f_1, \dots, f_p ) \in \bigcup_{ \bm{\delta} \in \{ + , -\}^p }  \bm{\mathbb{F}}_{\bm{\delta}}(\bm{\kappa}_{1, \bm{\delta}}, \bm{\kappa}_{2, \bm{\delta}}) $. 
\end{corollary}

\noindent For one-dimensional distributions, letting $f_j(t)=\alpha_j t$ and $p=2$, the KR model \eqref{kr:mult1d} simplifies to
\begin{align*}
T_{\bar\nu \rightarrow \nu_i} 
= 
T_{\varepsilon_i} \circ 
\Big(
\id 
+ \alpha_1 \big( T_{\bar\mu^1 \rightarrow \mu_i^1} - \id \big) 
+ \alpha_2 \big( T_{\bar\mu^2 \rightarrow \mu_i^2} - \id \big)
\Big),
\end{align*}
where $\alpha_1, \alpha_2 \in \mathbb{R}$.  The KR model is structurally different from the GOT model \cite{zhu2025geodesic, zhu2023atm}. In particular, the GOT model, without accounting for the ordering of predictors, can be written as
$
T_{\bar\nu \rightarrow \nu_i} 
=
T_{\varepsilon_i} 
\circ 
(\alpha_1 \odot T_{\bar\mu^1 \rightarrow \mu_i^1})
\circ
(\alpha_2 \odot T_{\bar\mu^2 \rightarrow \mu_i^2}),
$
where $
\alpha \odot T = \id + \alpha (T-\id), \text{ if } 0<\alpha<1,$
and $
\alpha \odot T = \id + \alpha (\id - T^{-1}), \text{ if } -1<\alpha<0.$
For $|\alpha|>1$, let $b=\lfloor |\alpha| \rfloor$ denote the integer part of $|\alpha|$, and set $a = |\alpha|-b$. The operator $\alpha \odot T$ is then defined by
\begin{align*}
\alpha \odot T(x) :=
\left\{
\begin{array}{ll}
(a \odot T) \circ 
\underbrace{T \circ T \circ \dots \circ T}_{b \text{ compositions of } T}(x), 
& \alpha > 1, \\[6pt]
(a \odot T^{-1}) \circ 
\underbrace{T^{-1} \circ T^{-1} \circ \dots \circ T^{-1}}_{b \text{ compositions of } T^{-1}}(x), 
& \alpha < -1 .
\end{array}
\right.
\end{align*}
Intuitively, the GOT model aggregates multiple predictors through compositions of transport maps, which are noncommutative, whereas KR adopts an additive structure. When the coefficient is negative, the ATM model uses the inverse transport map, while KR instead uses the opposite displacement direction.

\subsection{Mixed predictor}
The problem of predicting a distributional response $\nu_i$ with $q$ Euclidean predictors $\bm{X}_i = (X_i^1, \dots, X^{q}_i)^\top \in \mathbb{R}^q$ is of independent interest. To handle the Euclidean predictor $X_i^k$, the idea is to fit a $c$-concave function $\psi_k$ scaled uniformly by the centered covariate $Z_i^k = X_i^k - \E[ X_i^k ]$. We define the Kantorovich regression model with Euclidean predictors as
\begin{equation}\label{eq:euc}
    \E[ \phi_{\bar\nu \rightarrow \nu_i}(x) | \bm{X}_i ] = \sum_{k=1}^q Z_i^k \psi_k (x).
\end{equation}
where $\{\psi_1, \dots, \psi_q \}$ are $c$-concave functional parameters. 

Consider the case of a single Euclidean predictor $X_i$ ($q=1$).
The model reduces to
\[
\E[ \phi_{\bar\nu \rightarrow \nu_i}(x) | X_i ] = Z_i \psi (x), \text{ where }Z_i := X_i - \E[X].
\]
The effect of the Euclidean predictor on the distributional response is governed by the displacement field $-Z_i\nabla\psi(x)$, so that mass is transported locally along the direction $\nabla\psi(x)$ or its opposite depending on the sign of $Z_i$.
Specifically, when $Z_i>0$, mass is displaced in the direction $-\nabla\psi(x)$, whereas when $Z_i<0$, mass is displaced in the direction $\nabla\psi(x)$.
Thus, the Euclidean predictor modulates the response distribution by scaling and reversing the intrinsic transport direction encoded by the potential $\psi$. 

For notational convenience, we let $\bm{l} = (l^1, \dots, l^q)^\top \in \mathbb{R}^q$, $\bm{\rho} = (\rho^1, \dots, \rho^q)^T \in \mathbb{R}^q$ and $\bm{\mathbb{K}}(\bm{\rho}, \bm{\rho}) := \mathbb{K} (\rho^1, \rho^1 ) \times \dots \times \mathbb{K} (\rho^q, \rho^q)$. The following theorem characterizes a class of functions 
$(\psi_1,\dots,\psi_q)$, determined by the data-generating distribution 
of $\bm X_i$, that ensures the function 
$\sum_{k=1}^q Z_i^k \psi_k$ is $c$-concave.

\begin{theorem}\label{thm:euc}
Assuming that $| Z_i^k| \leq l^k $ holds almost surely for all $k$, then by choosing $\bm{\rho}$ to satisfy 
\begin{equation}\label{euc}
 \bm{l}^\top \bm{\rho}  \leq 1,
\end{equation}
we have that the function $\sum_{k=1}^q Z_i^k \psi_k $ is $c$-concave for any $(\psi_1, \dots,  \psi_q) \in \bm{\mathbb{K}} (\bm{\rho}, \bm{\rho} )$. 
\end{theorem}

Let $(\nu, \bm{\mu}, \bm{X}) \sim P_{\nu,  \bm{\mu}, \bm{X}}$, where $P_{\nu,  \bm{\mu}, \bm{X}}$ is the joint distribution of the random triplet $ (\nu, \bm{\mu}, \bm{X})$ taking values in the product space $ \mathcal{P}_{\mathrm{ac}}(\Omega) \times (\mathcal{P}_{\mathrm{ac}}(\Omega))^{p} \times \mathbb{R}^q$. The marginal distributions of $\nu, \bm{\mu}$ and $\bm{X}$ are denoted as $P_{\nu}$, $P_{\bm{\mu}}$ and $P_{\bm{X}}$ respectively. Given a sample $\{ \nu_i,  \bm{\mu}_i, \bm{X}_i \}_{i=1}^n $ that are i.i.d copies of $ (\nu, \bm{\mu}, \bm{X})$, we now extend the regression framework to accommodate $p$ distributional predictors $\bm{\mu}_i = (\mu^1_i, \dots, \mu_i^{p}) \in (\mathcal{P}_{\mathrm{ac}}(\Omega))^{p}$ and $q$ Euclidean predictors $\bm{X}_i = (X_i^1, \dots, X^{q}_i)^\top \in \mathbb{R}^q$ in predicting a distributional response $\nu_i \in \mathcal{P}_{\mathrm{ac}}(\Omega)$, and define the Kantorovich regression (KR) model with mixed predictors as
\begin{equation}\label{eq:mixed}
\E[ \phi_{\bar\nu \rightarrow \nu_i}(x) | \bm{\mu}_i, \bm{X}_i ]  = \sum_{j=1}^p \varphi_i^j (x) + \sum_{k=1}^q Z_i^k \psi_k (x), 
\end{equation}
where  $\varphi_i^j(x)  =  f_j \circledcirc \phi_{\bar\mu^j\to\mu_{i}^j} (x) - \E [f_j \circledcirc \phi_{\bar\mu^j\to\mu_{i}^j} (x)] $, and $\bm{f} = (f_1, \dots, f_p)^\top, \bm{\psi} =  (\psi_1, \dots, \psi_q)^\top$ are functional parameters. 

By setting $\bm{\rho}_{\bm{\delta}} = (\rho^1_{\bm{\delta}}, \dots, \rho^q_{\bm{\delta}})$, we mean that each component $\rho^k_{\bm{\delta}}$ depends on the whole $\bm{\delta}$. Similarly, the corollary below characterizes parameter spaces to ensure $c$-concavity of the predicted potential function.

\begin{corollary}\label{cor:mixed}
Assuming that $\phi_{\bar\mu^j\to\mu_{i}^j}\in\mathbb{K}(\gamma^{j}_-, \gamma^{j}_+ )$ and $| Z_i^k| \leq l^k $ hold almost surely for all $j$ and $k$, then for each $\bm{\delta} \in \{ + , -\}^p $, by choosing $ \bm{\kappa}_{1, \bm{\delta}}, \bm{\kappa}_{2, \bm{\delta}}, \bm{\rho}_{\bm{\delta}} $ to satisfy
\begin{equation}\label{eq:intrinsic_ineuqality}
(\bm{\gamma}_{\bm{\delta}}^\top + \bm{\lambda}^\top ) \bm{\kappa}_{1, \bm{\delta}}  + \bm{\eta}^\top \bm{\kappa}_{2, \bm{\delta}} + \bm{l}^\top \bm{\rho}_{\bm{\delta}}  \leq 1,
\end{equation}
the function $\sum_{j=1}^p \varphi_i^j + \sum_{k=1}^q Z_i^k \psi_k$ is $c$-concave for any $ (\bm{f}, \bm{\psi}) \in \cup_{ \bm{\delta} \in \{ + , -\}^p } \Theta_{\bm{\delta}}$, where $ \Theta_{\bm{\delta}} =  \bm{\mathbb{F}}_{\bm{\delta}}(\bm{\kappa}_{1, \bm{\delta}}, \bm{\kappa}_{2, \bm{\delta}}) \times \bm{\mathbb{K}}( \bm{\rho}_{\bm{\delta}}, \bm{\rho}_{\bm{\delta}} )$. 
\end{corollary}

\section{Asymptotic Theory}\label{sec:asym}

Given functions $\phi : \Omega \rightarrow \mathbb{R}$ and $T:\Omega \rightarrow \Omega$, we denote their $L^2(\bar\nu)$-norm respectively as
$
\| \phi \|_{L^2(\bar\nu)} := (\int (\phi(x))^2 \; \rd \bar\nu(x))^{1/2} \text{ and } \| T \|_{L^2(\bar\nu)} := (\int \| T(x) \|^2 \; \rd \bar\nu(x))^{1/2}$.  With a slight abuse of notation, we use the same symbol $\nu$ to denote both a measure and its density when it exists. The sup-norm of a density $\nu$ is defined by $\|\nu\|_{\infty} := \sup_{x \in \mathrm{sp}(\nu)} \nu(x)$.

Given a sample $\{ \nu_i, \bm{\mu}_i, \bm{X}_i\}_{i=1}^n$, the intercept term $\E [f_j \circledcirc \phi_{\bar\mu^j\to\mu_{i}^j}(x)]$ and $\E[X_i^k]$ 
are estimated by their sample averages 
$\frac{1}{n} \sum_{i=1}^n f_j \circledcirc \phi_{\bar\mu^j_n \to\mu_{i}^j} (x)$ 
and $\frac{1}{n} \sum_{i=1}^n X_i^k$, respectively. The functional parameters are estimated by minimizing the empirical Sobolev loss 
\begin{align*}
    \widehat{L}(\bm{f}, \bm{\psi}) := & \frac{1}{n} \sum_{i=1}^n \Big\|  \phi_{\bar\nu_n \rightarrow \nu_i} -   \Big(\sum_{j=1}^p \widehat{\varphi}_i^j + \sum_{k=1}^q \widehat{Z}_i^k \psi_k \Big)  \Big\|_{\dot{H}^1(\bar\nu_n)}^2 \\
    =& \frac{1}{n} \sum_{i=1}^n \Big\| \nabla \phi_{\bar\nu_n \rightarrow \nu_i} - \nabla  \Big(\sum_{j=1}^p \widehat{\varphi}_i^j + \sum_{k=1}^q \widehat{Z}_i^k \psi_k \Big)  \Big\|_{L^2(\bar\nu_n)}^2,
\end{align*}
where $ \widehat{\varphi}_i^j = f_j \circledcirc \phi_{\bar\mu^j_n \to\mu_{i}^j}(x)  -\frac{1}{n} \sum_{l=1}^n f_j \circledcirc \phi_{\bar\mu^j_n \to\mu_{l}^j} (x)$ and $\widehat{Z}_i^k = X_i^k - \frac{1}{n} \sum_{l=1}^n X^k_l$. Its population counterpart is defined by
\[
L(\bm{f}, \bm{\psi})
:=
\E \Big[ \Big\|
\nabla \phi_{\bar\nu \rightarrow \nu_1} - \nabla \Big( \sum_{j=1}^p \varphi_1^j 
+ \sum_{k=1}^q Z_1^k \psi_k 
\Big) \Big\|_{L^2(\bar\nu)}^2
\Big].
\]
If $\nu_1$ is generated from model \eqref{eq:mixed} with true parameter 
$(\check{\bm f}, \check{\bm \psi})$, then $(\check{\bm f}, \check{\bm \psi})$ is a minimizing pair of 
$L(\bm f,\bm \psi)$. The following theorem establishes the uniform convergence of 
$\widehat{L}$ to $L$ over $\cup_{ \bm{\delta} \in \{ + , -\}^p } \Theta_{\bm{\delta}}$ with an explicit convergence rate. We note that the inequality \eqref{eq:intrinsic_ineuqality} is used to enforce intrinsic structure of the model, and is not needed here.

\begin{theorem}\label{StatConvrgence}
Suppose the following regularity assumptions: (1) $\Omega$ is convex and that there exists a constant $C_1>0$ such that
$\inf_{x\in\Omega}\nu_i(x)\ge C_1$ and
$\inf_{x\in\Omega}\mu_i^j(x)\ge C_1$
almost surely for all $i$ and $j$;
(2) $\|\bar\nu_n\|_{\infty}=O_p(1)$,
$\|\bar\mu_n^j\|_{\infty}=O_p(1)$ for all $j$,
and $\|\bar\nu\|_{\infty}<\infty$.
(3) $|Z_i^k|\le l^k$ almost surely for all $i$ and $k$, and that,
for each $\bm{\delta}\in\{+,-\}^p$, the vectors
$\bm{\kappa}_{1,\bm{\delta}}$,
$\bm{\kappa}_{2,\bm{\delta}}$, and
$\bm{\rho}_{\bm{\delta}}$
are bounded; (4) $\phi_{\bar\mu^j\to\mu_i^j},
\phi_{\bar\nu\to\nu_i}
\in\mathbb{K}(\gamma_-,\gamma_+)$ almost surely
and $\gamma_+<1$ if $d>1$ for all $i$ and $j$. We have
\begin{align*}
\sup_{(\bm{f},\bm{\psi})
\in
\cup_{\bm{\delta}\in\{+,-\}^p}\Theta_{\bm{\delta}}}
\left|
\widehat{L}(\bm{f},\bm{\psi})
-
L(\bm{f},\bm{\psi})
\right|
= \left\{
\begin{array}{cc}
O_p(n^{-1/2}),  & d=1; \\
O_p(n^{-1/8}),  & d>1.
\end{array} \right.
\end{align*}
\end{theorem}

The regularity conditions assumed in the above theorem are standard in the optimal transport literature and are commonly imposed in the study of both computational algorithms \cite{paty2020regularity, kim2025optimal} and statistical convergence properties \cite{10.1214/20-AOS1997, manole2024plugin, li2026smoothed}, among others. When $d>1$, the convergence results are based on the quantitative stability bound of optimal transport  $ \| T_{\bar{\nu} \to \nu} - T_{\bar{\nu}' \to \nu} \|_{L^2(\Omega)} \lesssim (W_1(\bar\nu, \bar\nu'))^{1/2}$ in \cite{berman2021convergence}, and recently developed convergence rate of Wasserstein barycenter $W_2(\bar\nu_n, \bar\nu) = O_p(n^{-1/4})$ \cite{leon2026wasserstein}.

The population loss $L(\bm f,\bm\psi)$ and the empirical loss $\widehat L(\bm f,\bm\psi)$ are convex in $(\bm f,\bm\psi)$. For each fixed $\bm\delta$, the parameter space $\Theta_{\bm\delta}$ is also convex. Therefore, in practice, a minimizer of $\widehat L$ over each $\Theta_{\bm\delta}$ can be obtained using projected gradient descent. Additional details on optimization algorithms are postponed to Section B in Appendix. A global minimizer over $\bigcup_{\bm\delta\in{+,-}^p}\Theta_{\bm\delta}$ is then obtained by selecting, among the minimizers over the individual sets $\Theta_{\bm\delta}$,  anyone that attains the smallest value of $\widehat L$.

Let $\check{\theta}= (\check{\bm f}, \check{\bm \psi})$ be the true parameter and ${\widehat\theta}_n = ( \widehat{\bm f}, \widehat{\bm \psi} )$ be the minimizer of $\widehat L$ in $\cup_{ \bm{\delta} \in \{ + , -\}^p}\Theta_{\bm\delta}$, we derive consistency in terms of excess risk $d_h(\widehat{\theta}_n, \check{\theta})$, where $d_h^2(\theta_1,\theta_2) 
:=\E[
\|
\nabla h_{\theta_1}
-
\nabla h_{\theta_2}
\|_{L^2(\bar\nu)}^2
]$ and $
h_\theta(x) 
=
\sum_{j=1}^p  \varphi_i^j(x)
+
\sum_{k=1}^q Z_i^k  \psi_k (x)$ with $\theta = (\bm f,\bm\psi)$. 

\begin{theorem}\label{lem:param_consistency}
Under the assumptions of Theorem~\ref{StatConvrgence}, we have
$d_h^2(\widehat{\theta}_n,\check{\theta})=O_p(n^{-1/2})$ if $d=1$, and
$d_h^2(\widehat{\theta}_n,\check{\theta})=O_p(n^{-1/8})$ if $d>1$.
\end{theorem}

\section{Numerical Studies}\label{sec:exp}

\subsection{1D synthetic data}
We examine the proposed model on a one-dimensional synthetic data consisting of two distributional predictors and one distributional response. The performance is compared with two benchmark methods: Wasserstein regression (WR) [\cite{chen2023wasserstein}], and distribution-on-distribution regression (DoD) [\cite{ghod:21}]. For fair comparison, we generate the synthetic distributions directly through their quantile functions, rather than from any of the three regression models. 

In particular, we first defined the basis functions $
b_k(t) = \frac{\sin(k\pi t)}{k\pi}, t \in [0,1]$, which introduce smooth perturbations of the uniform quantile function $Q(t)=t$ at different frequencies. For each observation $i$, we independently generated $A_i,B_i,C_i,D_i \sim \operatorname{Unif}[-0.25,0.25]$ and constructed the quantile functions of the two distributional predictors $\mu_i^1$, $\mu_i^2$ as $
Q_{\mu_i^1}(t) = t + A_i b_1(t) + B_i b_2(t), Q_{\mu_i^2}(t) = t + C_i b_1(t) + D_i b_3(t)$ respectively.
Thus, the random coefficients induce smooth subject-specific variations in the shapes of the two predictor distributions. The quantile function of the distributional response $\nu_i$ was generated as
\[
Q_{\nu_i}(t)
=
0.95\left\{
0.6 h_{0.8}\left(Q_{\mu_i^1}(t)\right)
+
0.4 h_{-0.8}\left(Q_{\mu_i^2}(t)\right)
\right\}
+
0.05\left\{
t + \eta_i b_4(t)
\right\},
\]
where $
h_\gamma(x) = x + \gamma x(1-x)(2x-1)$ is a nonlinear monotone transformation that introduces nonlinear distortions of the predictor quantile functions, with the sign of $\gamma$ controlling the direction of the distortion. The first component represents the systematic relationship between the response and the two distributional predictors, with respective weights $0.6$ and $0.4$. The second component is centered around the uniform quantile function $t$ and introduces additional smooth random variation through $\eta_i b_4(t)$, where $\eta_i \sim \operatorname{Unif}[-0.15,0.15]$. The higher-frequency basis function $b_4$ therefore captures variation in the response that is independent of the predictors. Finally, the quantile functions $Q_{\mu_i^1}$, $Q_{\mu_i^2}$, and $Q_{\nu_i}$ were converted into the corresponding densities $\mu_i^1$, $\mu_i^2$, and $\nu_i$ on $[0,1]$, respectively.

We performed 200 Monte Carlo repetitions. In each repetition, we generated a sample of 60 triplets $\{(\mu_i^1,\mu_i^2,\nu_i)\}_{i=1}^{60}$ and split them into 50 training observations and 10 test observations. Each model was fitted using the training set $\{(\mu_i^1,\mu_i^2,\nu_i)\}_{i=1}^{50}$, and its predictive performance was evaluated on the test set $\{(\mu_i^1,\mu_i^2,\nu_i)\}_{i=51}^{60}$ using the mean prediction error $
\frac{1}{10}\sum_{i=51}^{60} W_2^2(\nu_i,\widehat{\nu}_i),$
where $\widehat{\nu}_i$ denotes the response density predicted by each model. We report the average mean prediction error and its standard deviation across the 200 Monte Carlo repetitions. Since DoD regression cannot accommodate multiple distributional predictors, we fitted two DoD regression models using $\mu_i^1$ and $\mu_i^2$ as predictors, respectively, and reported the corresponding prediction errors for both models. Table \ref{prediction_errors:1Dsynthetic} shows that Kantorovich regression attains the smallest average and the smallest standard deviation for mean prediction errors.

\begin{table}[ht]
\centering
\caption{Average and standard deviation of the mean prediction errors.}
\label{prediction_errors:1Dsynthetic}
\begin{tabular}{lcccc}
\hline
 & KR & WR & DoD with $\mu_i^1$ & DoD with $\mu_i^2$ \\
\hline
prediction error
& $\bf{1.904\times 10^{-5}}$
& $2.682\times 10^{-5}$
& $2.552\times 10^{-4}$
& $1.163\times 10^{-3}$ \\
standard deviation
& $\bf{5.803\times 10^{-6}}$
& $1.325\times 10^{-5}$
& $7.604\times 10^{-5}$
& $3.849\times 10^{-4}$ \\
\hline
\end{tabular}
\end{table}

\subsection{1D real data}
In housing market, sale prices are realized after search, bargaining, and financing, their distribution ($\nu$) should be a deformation of the listing distribution ($\mu$), encoding sellers' initial offers. The mortgage rate ($X$) shapes buyers' purchasing power and bidding intensity: higher rates tighten affordability and shift mass in $\nu$ toward lower prices, whereas lower rates ease constraints and push the distribution rightward. In light of these mechanisms, we predict the monthly density \(\nu\) of median sale prices  for single-family residences (SFRs) across U.S. metropolitan areas over March 2018-December 2024, using predictors \((\mu_i, X_i)\) available between January 2018 and October 2024. Here, we denote \(\mu_i\) and \(X_i\) as the listing-price density for SFR and the mortgage rate observed two months prior to \(\nu_i\) (with \(X_i\) measured in \%). We model the relationship between $\nu_i$ and $(\mu_i, X_i)$ by the proposed Kantorovich regression with mixed predictors:
$\E [T_{\bar\nu \to \nu_i} | \mu_i, Z_i] =  T_{\check{f} \circledcirc \phi_{\bar\mu \rightarrow \mu_i} -\E[\check{f} \circledcirc \phi_{\bar\mu \rightarrow \mu_i}] + Z_i \check{\psi}}$.

\begin{figure}
\centering
\includegraphics[width=0.4\linewidth]{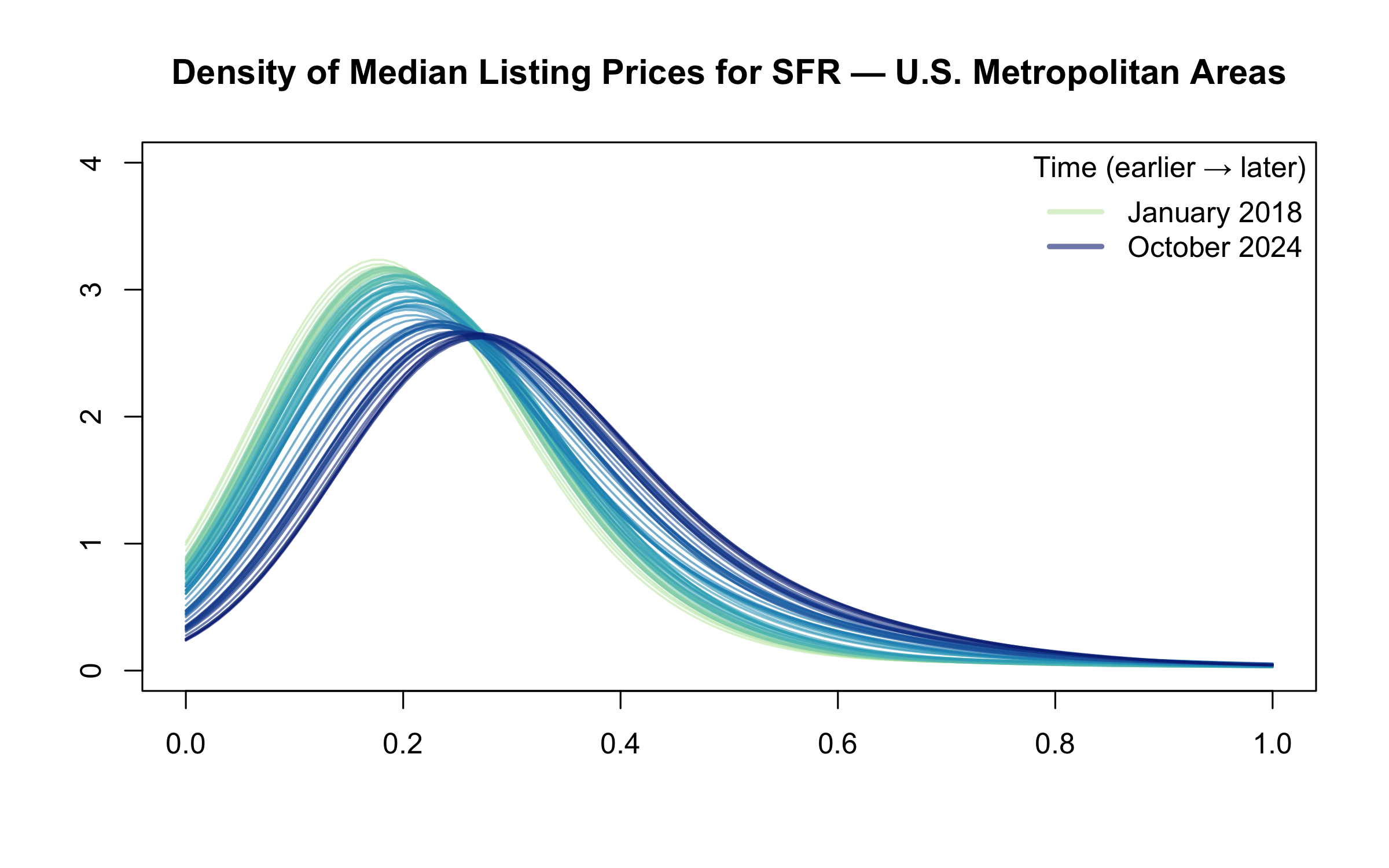}
\centering
\includegraphics[width=0.4\linewidth]{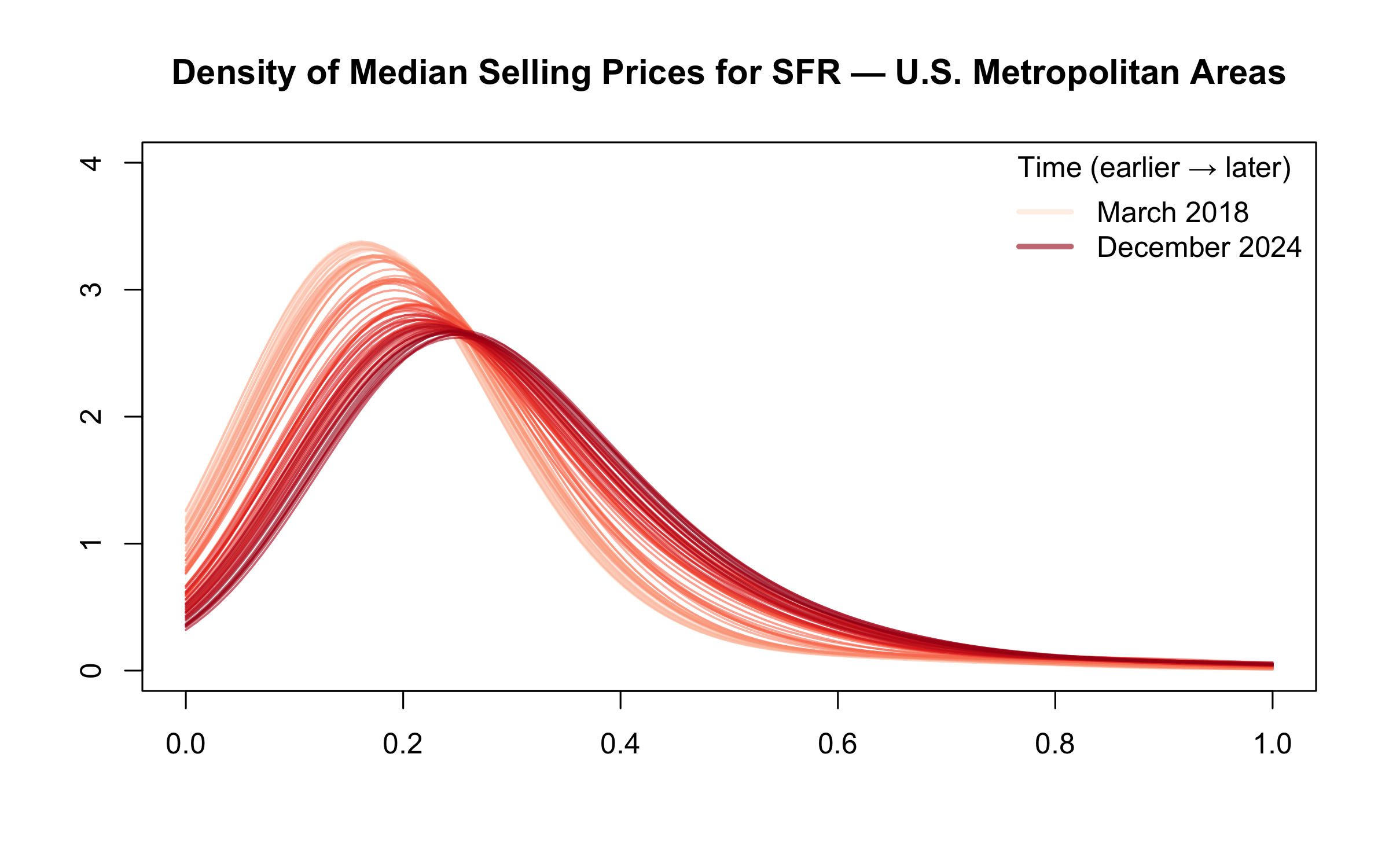}
\includegraphics[width=0.4\linewidth]{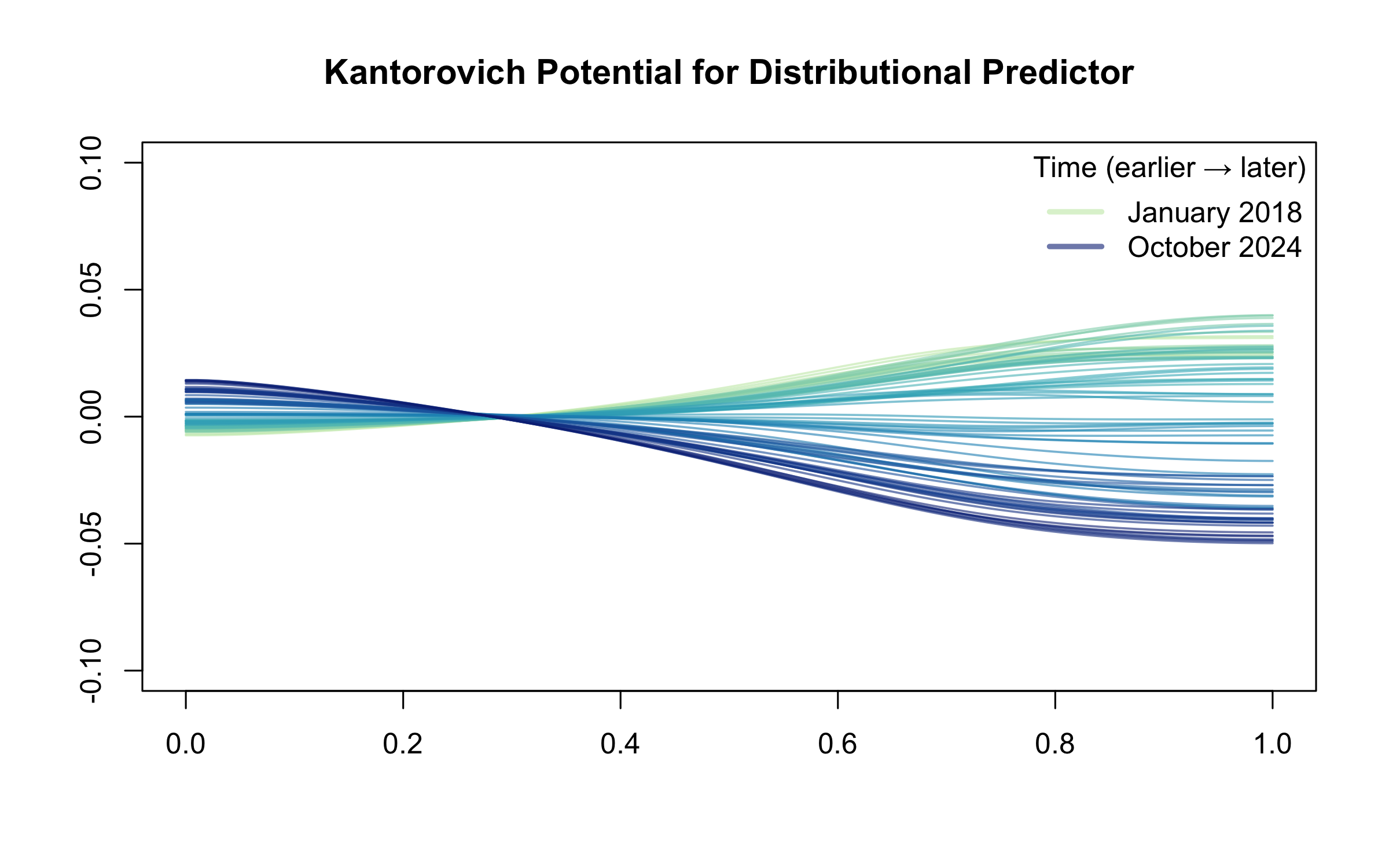}
\centering
\includegraphics[width=0.4\linewidth]{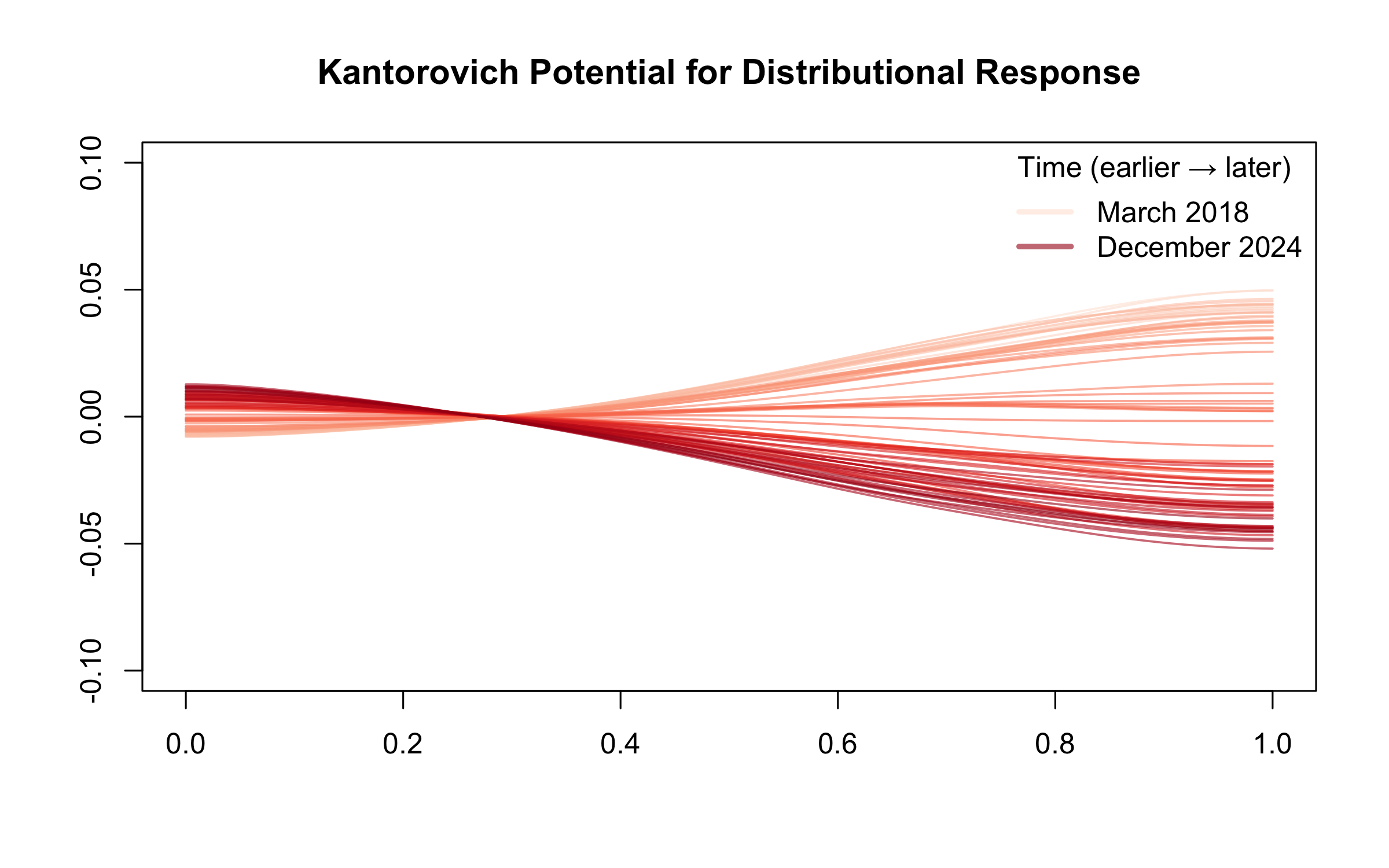}
\includegraphics[width=0.4\linewidth]{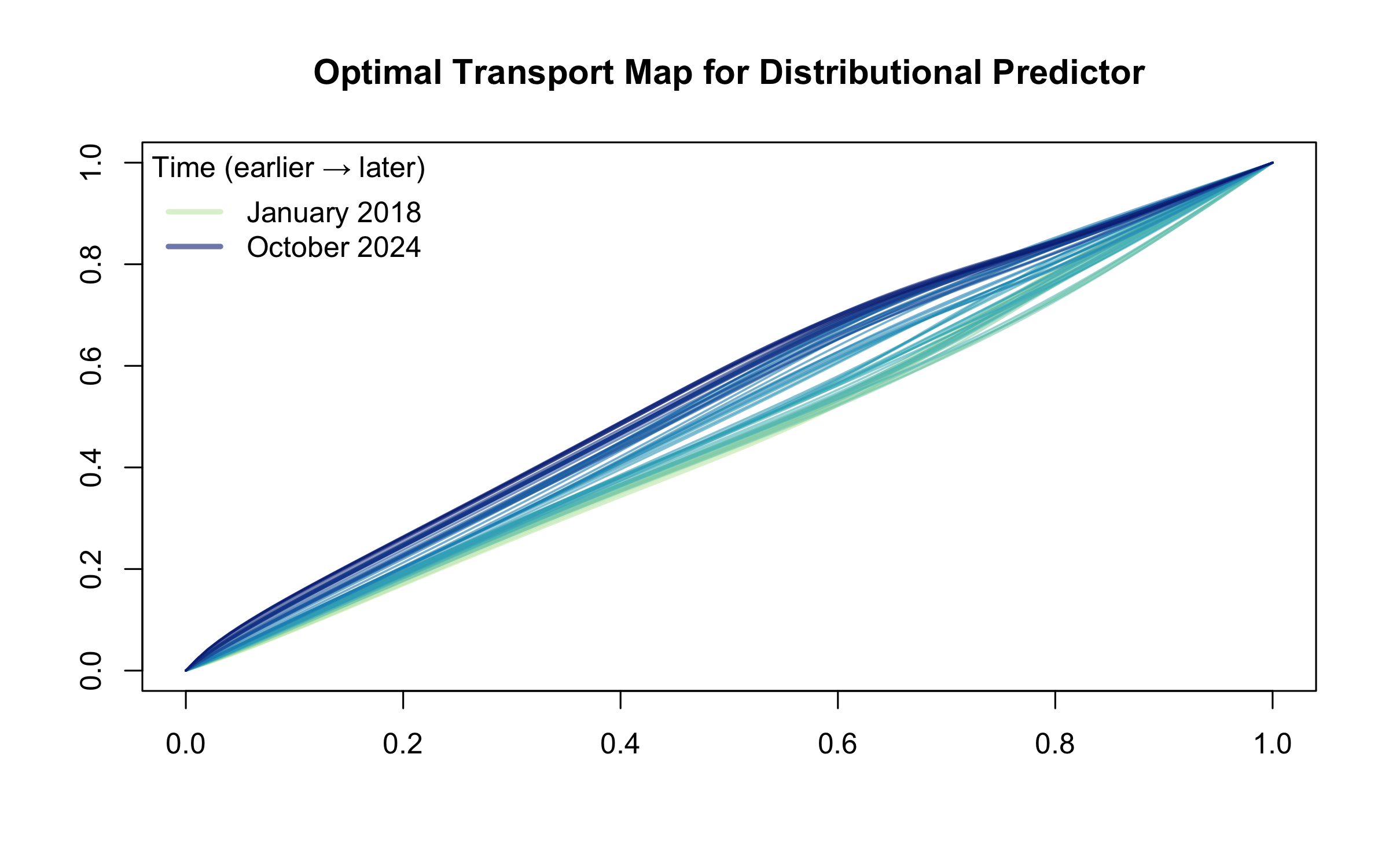}
\centering
\includegraphics[width=0.4\linewidth]{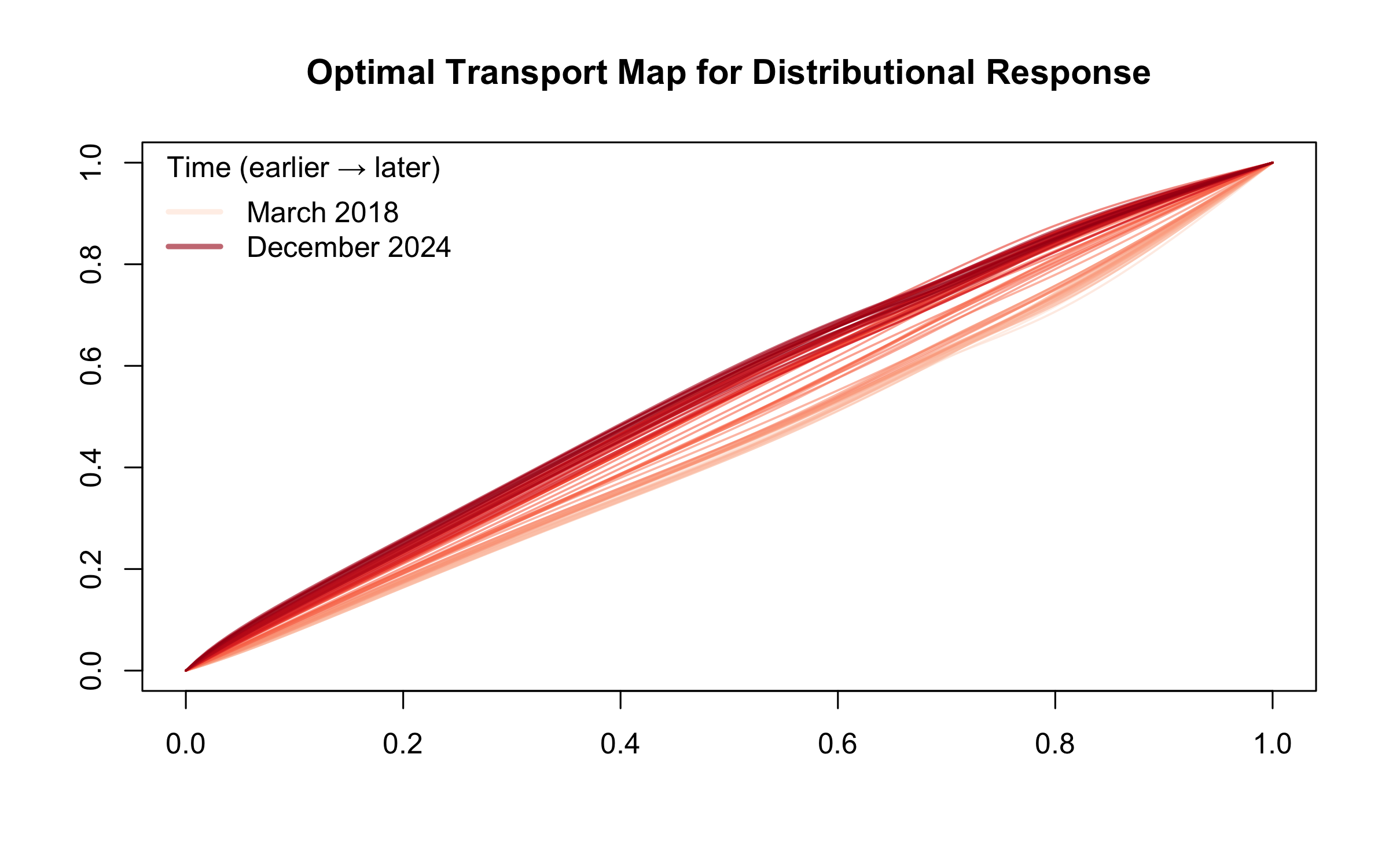}
\includegraphics[width=0.8\linewidth]{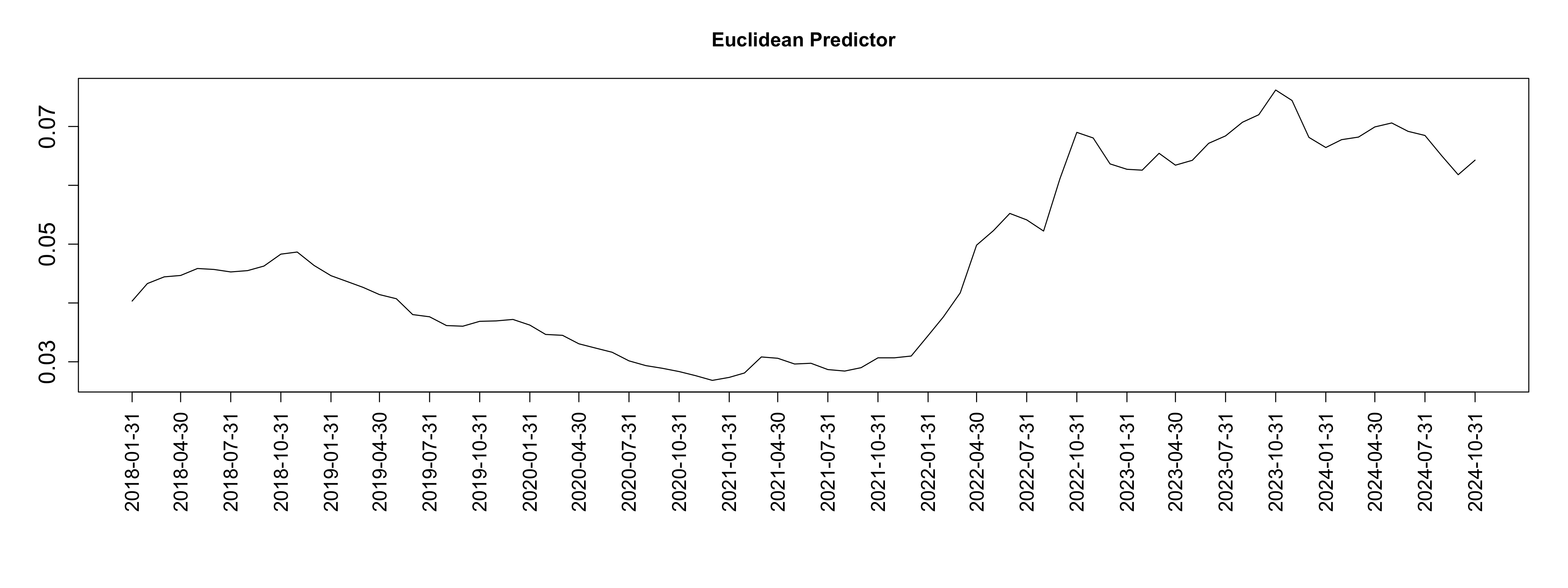}
\caption{(Top left) Observed distributional predictor, (Top right) observed distributional response, (Upper left) observed Kantorovich potential for distributional predictor, (Upper right) observed Kantorovich potential for the distributional response, (Lower left) observed optimal transport map for distributional predictor, (Lower right) observed optimal transport map for the distributional response, and (Bottom) Euclidean Predictor.}
\label{fig:data}
\end{figure}

Monthly median sale and listing price series for SFR for all metropolitan areas in the US are available from Zillow(\url{https://www.zillow.com/research/data/}). Before estimating the densities, we divided all prices by $\$1$M and set the supports of $\left\{\mu_i\right\}_{i=1}^{82}$ and $\left\{\nu_i\right\}_{i=1}^{82}$ to be $\left[0,1\right]$. We estimated the densities through kernel density estimation with bandwidth 0.1. For the Euclidean predictor, we obtained the FRED's 30-year fixed mortgage series MORTGAGE30US from Python package \texttt{pandas datareader}, a weekly national average commitment rate for conforming conventional purchase loans. Weekly values are averaged to monthly to align with our listing and sale price densities. The distributional responses and predictors, together with the corresponding Kantorovich potentials and optimal transports, as well as the Euclidean predictors, are visualized in Figure \ref{fig:data}. 

\begin{figure}[t]
    \centering
\begin{tikzpicture}
  \begin{groupplot}[
      group style={group size=2 by 1, horizontal sep=8mm, vertical sep=6mm, group name=G}
  ]

  \nextgroupplot[fxplot3,tick label style={font=\scriptsize},
  tick style={draw=none},tight ticks,legend style={
    at={(0.02,0.98)}, anchor=north west,
    font=\tiny, inner sep=1pt, nodes={inner sep=1pt},
    row sep=3pt, column sep=2pt,
    /tikz/line width=0.4pt
  }]
    \addplot[black] table[col sep=comma, header=false, x index=0, y index=1]{Asset/realdata/1D/csv/g.csv};
    \legend{$\widehat{f}'$}

  \nextgroupplot[psiplot,tick label style={font=\scriptsize},tick style={draw=none},tight ticks,legend style={
    at={(0.02,0.98)}, anchor=north west,
    font=\tiny, inner sep=1pt, nodes={inner sep=1pt},
    row sep=3pt, column sep=2pt,
    /tikz/line width=0.4pt
  }]
    \addplot[black] table[col sep=comma, header=false, x index=0
    , y index=1]{Asset/realdata/1D/csv/minus_grad_psi.csv};
    \addplot[domain=0:1, samples=2, dotted, line width=0.3pt]{0};
    \legend{$-{\widehat\psi}'$, $y=0$}
  \end{groupplot}
\end{tikzpicture}
\caption{(Left) Estimated derivative \(\widehat f'\) associated with the distributional predictor; (Right) Negative derivative \(-\widehat\psi'\) of estimated functional parameter $\hat\psi$ associated with the Euclidean predictor.}
\label{fig:parameter1D}
\end{figure}
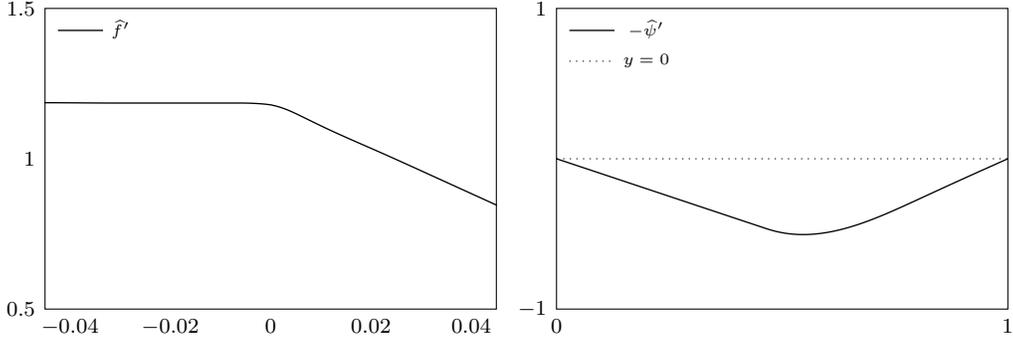

To demonstrate the interpretability of the proposed model, we first fit it to the entire dataset. The fitted parameter $\widehat{f}'$ is shown in the left panel of 
Figure \ref{fig:parameter1D}. We observe that $f'(t) \approx 1.2$ for 
$t \le 0$, after which it decreases approximately linearly and falls 
below $1$ once $t > 0.02$. Thus, the KR model amplifies displacements for small values of $\phi$ and attenuates them for larger values. When the listing price distribution lies to the right of $\bar\mu$, 
$\nabla \phi_{\bar\mu \to \mu} < 0$ and $\phi_{\bar\mu \to \mu}$ is decreasing. In this case, smaller values of $\phi$ correspond to the upper tail of the listing distribution (i.e., higher-priced homes). Since $f'(\phi)$ is larger in this region, the KR model amplifies 
the displacement from the barycenter in the upper tail of the sale distribution, while displacement in the lower tail is decreased. Conversely, when the listing distribution lies to the left of 
$\bar\mu$, $\nabla \phi_{\bar\mu \to \mu} > 0$ and $\phi_{\bar\mu \to \mu}$ 
is increasing. Smaller values of $\phi$ then correspond to the lower 
tail of the listing distribution (i.e., lower-priced homes). The KR model amplifies the 
displacement from the barycenter in the lower tail of the sale 
distribution, while displacement in the upper tail is attenuated. In economic languages, the model suggests that when the listing distribution lies to the right 
of the barycenter (i.e., the market is relatively strong/confident), the higher-price segment exhibits a larger deviation from the market barycenter in the sale 
distribution than in the listing distribution. Conversely, when the listing distribution 
lies to the left of the barycenter (i.e., the market is relatively weak), 
the lower-price segment exhibits a larger deviation.

On the other hand, the estimated functional parameter $\widehat \psi$ associated with the Euclidean predictor $X_i$, shown in the right panel of Figure \ref{fig:parameter1D}, satisfies $- \widehat\psi'(x)\le 0$ for any $x \in [0,1]$. Recall that $-\psi'(x)$ represents the additional displacement between an individual distribution and the barycenter. Hence, the fact that $-\widehat{\psi}'(x)$ is uniformly negative indicates a leftward deformation that strengthens as $X_i$ increases. This pattern accords with market intuition: when mortgage rates are higher, sale-price distributions shift toward lower values. 

To evaluate the predictive performance, we considered a rolling-window prediction scheme using the same data, and compared with Wasserstein regression (WR) [\cite{chen2023wasserstein}] and distribution-on-distribution regression (DoD) [\cite{ghod:21}]. Since WR and DoD do not accommodate Euclidean predictors, the Euclidean predictor is omitted for these methods in this example. In the first window, each model was fitted using the 50 predictor–response pairs $\{(\mu_i,\nu_i)\}_{i=1}^{50}$, and $\nu_{51}$ was predicted from $\mu_{51}$. The training window was then shifted forward by one month at a time until reaching the final window, in which the models were trained using the observations $\{(\mu_i,\nu_i)\}_{i=32}^{81}$, and $\nu_{82}$ was predicted from $\mu_{82}$. For each window, we computed the squared Wasserstein distance between the observed and predicted response distributions as the prediction error. Table~\ref{prediction_errors:1Dreal} shows that KR achieves both the smallest average prediction error and the smallest standard deviation across 32 rolling windows.

\begin{table}[ht] \centering \caption{Comparison of average prediction errors with standard deviations.} \label{prediction_errors:1Dreal} \begin{tabular}{lccc} \hline & KR & WR & DoD \\ \hline Average prediction error & $\bf{9.464\times 10^{-5}}$ & $1.343\times 10^{-4}$ & $1.568\times 10^{-4}$ \\ Standard deviation & $\bf{1.135\times 10^{-4}}$ & $1.597\times 10^{-4}$ & $1.237\times 10^{-4}$ \\ \hline \end{tabular} 
\end{table}

\subsection{2D real data}
In this second real-data example, we demonstrate that Kantorovich regression can be applied to bivariate distributional predictors and responses. We use the joint distributions of daily minimum and maximum temperatures during the summer months (June–August) of 2000 and 2020, measured at 50 airports in the US. The daily minimum and maximum temperature data is available from \url{https://www.ncdc.noaa.gov/cdo-web/datasets}. We estimate the densities on the domain $[20, 120] \times [40, 140]$, where the first and second coordinates correspond to daily minimum and maximum temperatures, respectively, discretized on a $200 \times 200$ grid. The densities are obtained by smoothing two-dimensional histograms using a Gaussian filter. Using the bivariate distributions of daily minimum and maximum temperatures in summer 2000 as predictors, we aim to predict the corresponding temperature distributions in summer 2020 via model \ref{eq:single}. 

\begin{wrapfigure}{r}{0.4\textwidth}
\centering
\begin{tikzpicture}
  \begin{groupplot}[
      group style={group size=1 by 1, horizontal sep=16mm, vertical sep=12mm, group name=G}
  ]

  \nextgroupplot[
      fxplot6,
      tick label style={font=\scriptsize},
      tick style={draw=none},
      tight ticks,
      legend style={
        at={(0.02,0.98)}, anchor=north west,
        font=\tiny, inner sep=1pt, nodes={inner sep=1pt},
        row sep=3pt, column sep=2pt,
        /tikz/line width=0.4pt
      }
  ]
    \addplot[black] table[col sep=comma, header=false, x index=0, y index=1]{Asset/realdata/2D/csv/result2d.csv};
    \legend{$\widehat f'$}

  \end{groupplot}
\end{tikzpicture}
\caption{Plot of estimated \(\widehat{f}'\).}
\label{2D:sim}
\end{wrapfigure}

\begin{figure}[!htbp]
\centering
\begin{tikzpicture}[scale=0.9]

\pgfplotsset{
    densityaxis/.style={
        width=0.2\textwidth,
        height=0.2\textwidth,
        xmin=20, xmax=120,
        ymin=40, ymax=140,
        axis equal image,
        enlargelimits=false,
        scale only axis,
        axis on top,
        axis line style={draw=none},
        xtick={20,40,60,80,100,120},
        ytick={40,60,80,100,120,140},
        tick label style={font=\tiny},
        tick style={draw=none},
        tight ticks
    },
    emptyaxis/.style={
        width=0.2\textwidth,
        height=0.2\textwidth,
        xmin=20, xmax=120,
        ymin=40, ymax=140,
        hide axis
    }
}

\begin{groupplot}[
    group style={
        group size=4 by 3,
        horizontal sep=8mm,
        vertical sep=6mm,
        group name=G
    }
]


\nextgroupplot[emptyaxis]
{}

\nextgroupplot[densityaxis]
\addplot graphics[
    xmin=20, xmax=120,
    ymin=40, ymax=140
]{
    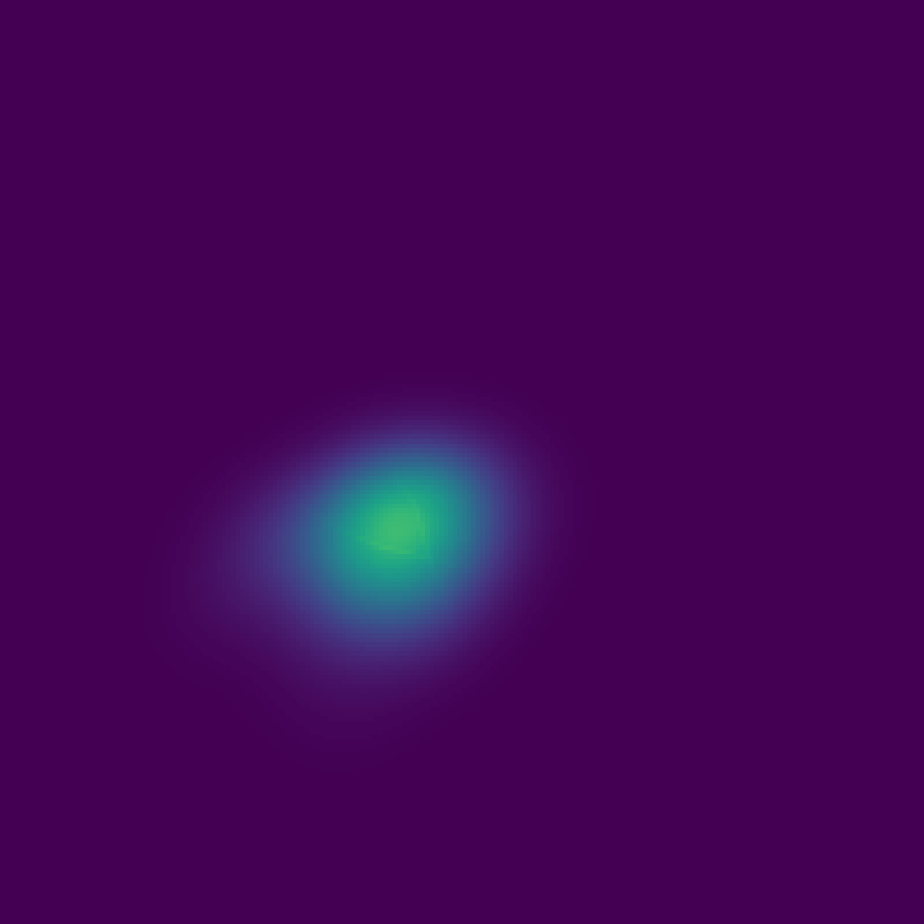
};

\nextgroupplot[densityaxis]
\addplot graphics[
    xmin=20, xmax=120,
    ymin=40, ymax=140
]{
    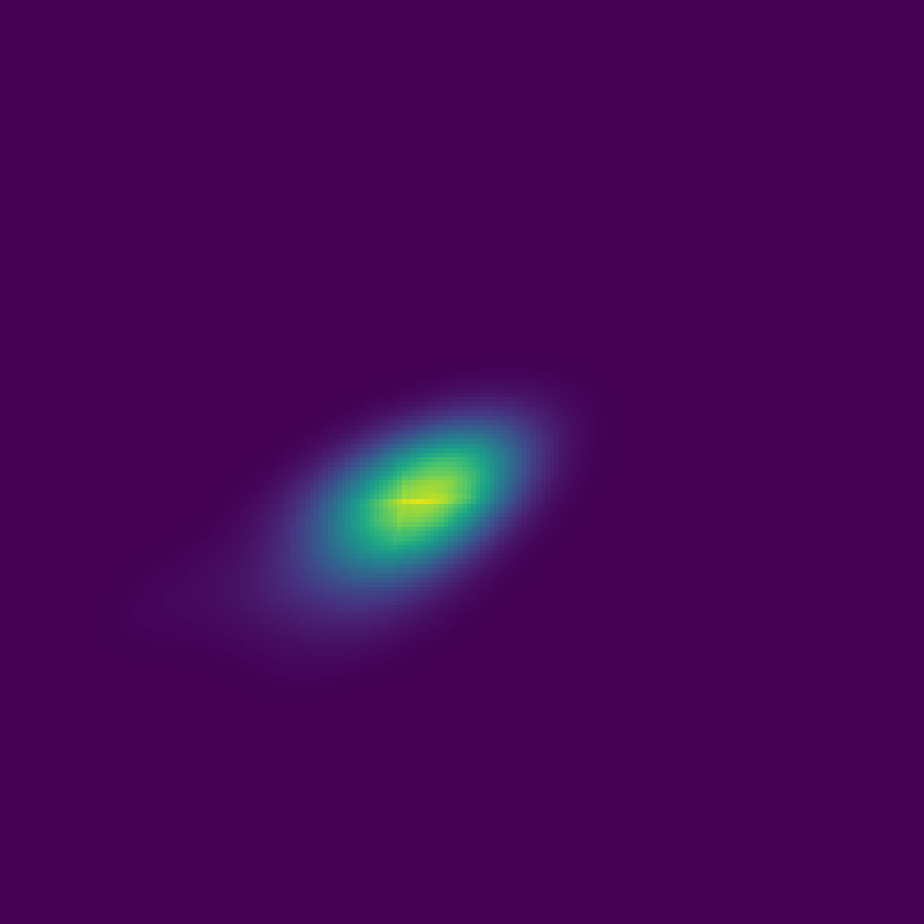
};

\nextgroupplot[densityaxis]
\addplot graphics[
    xmin=20, xmax=120,
    ymin=40, ymax=140
]{
    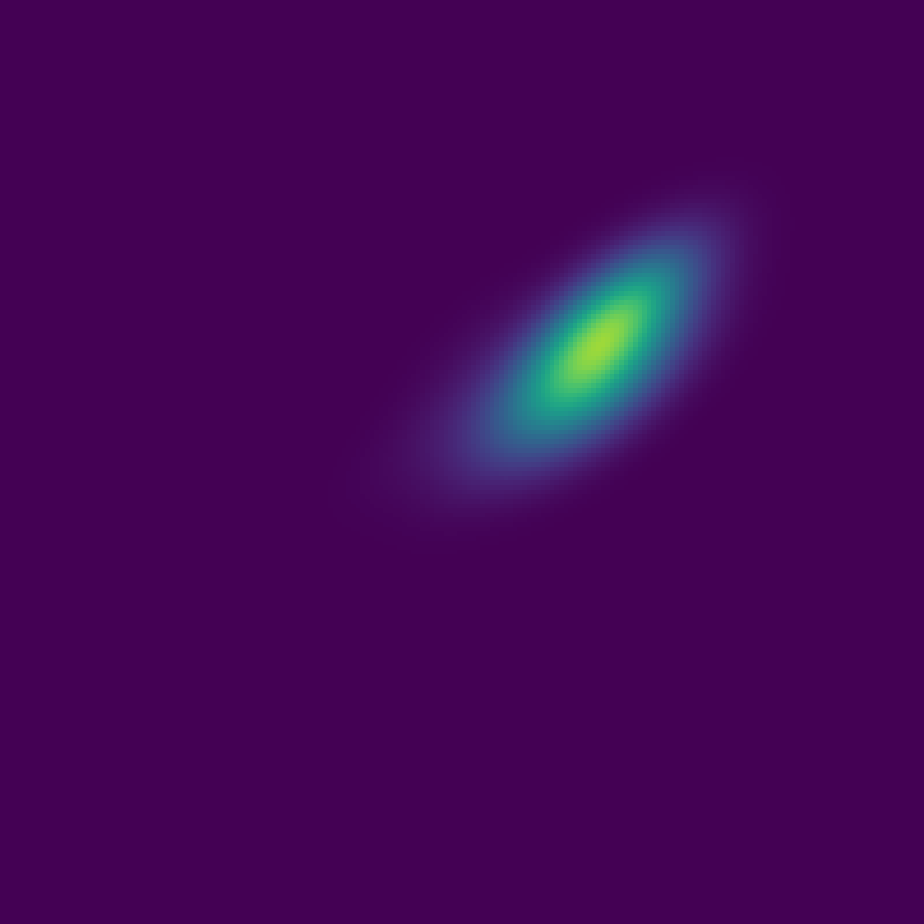
};


\nextgroupplot[densityaxis]
\addplot graphics[
    xmin=20, xmax=120,
    ymin=40, ymax=140
]{
    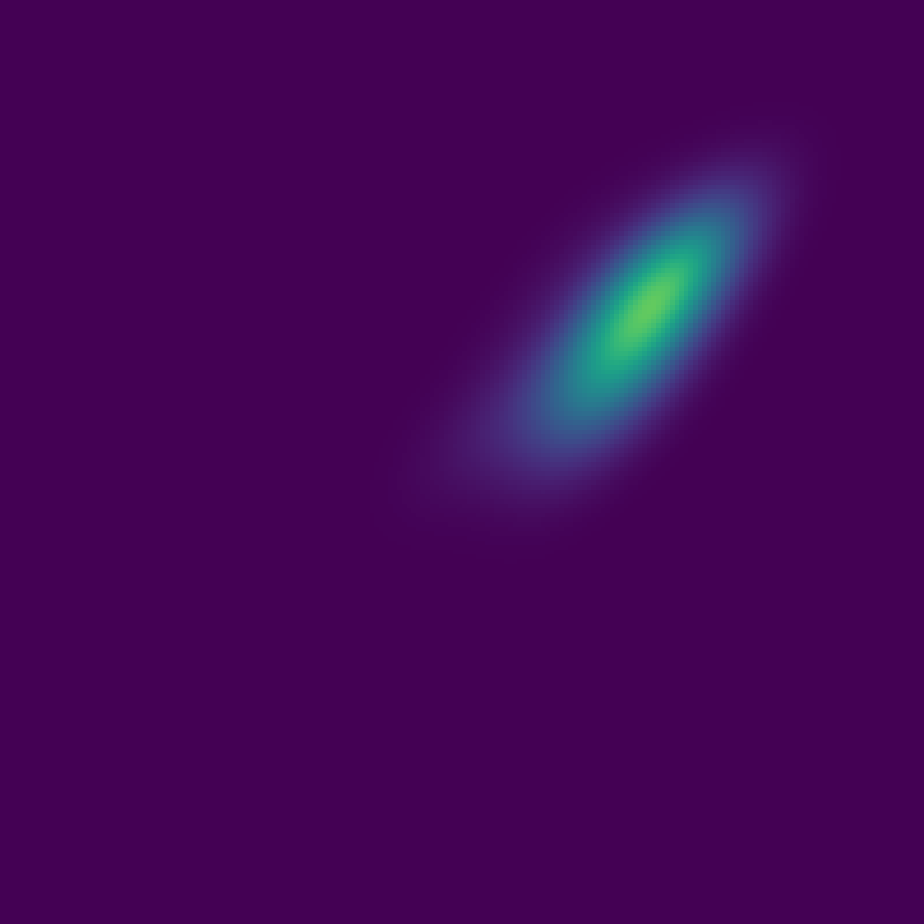
};

\nextgroupplot[densityaxis]
\addplot graphics[
    xmin=20, xmax=120,
    ymin=40, ymax=140
]{
    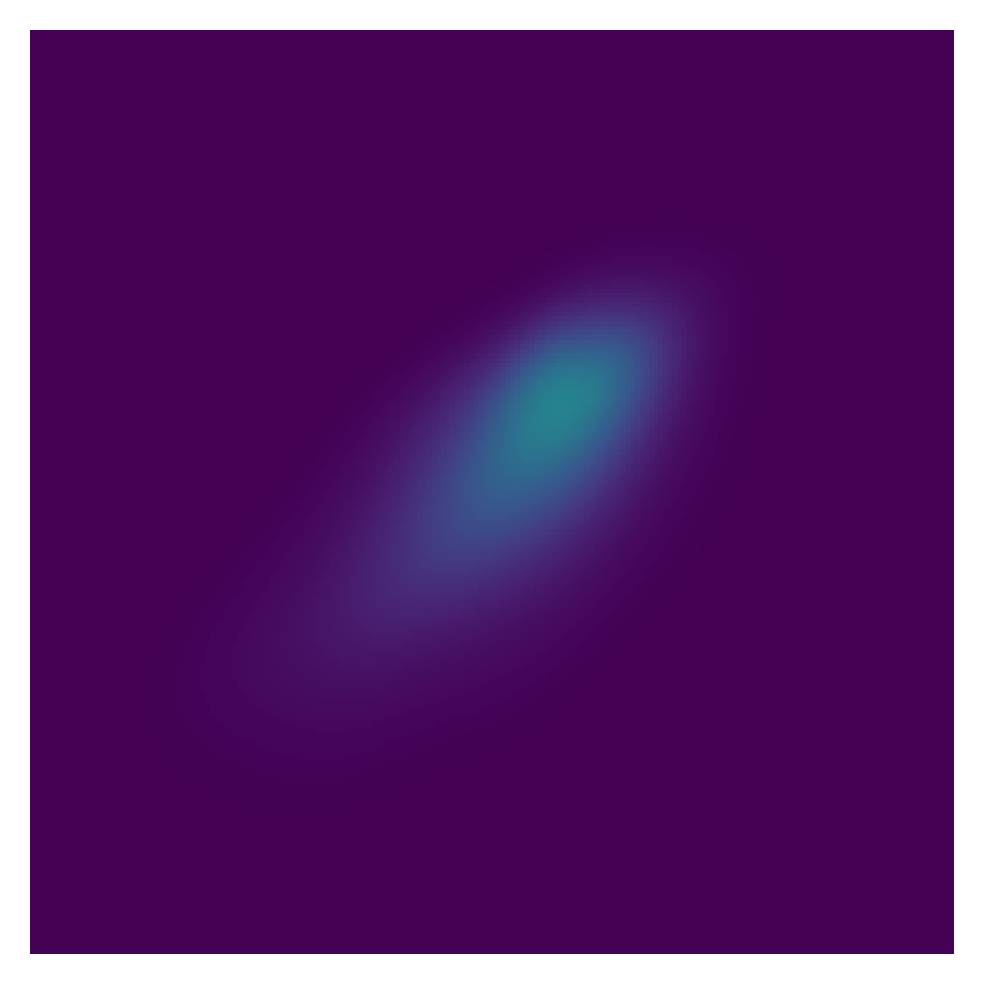
};

\nextgroupplot[densityaxis]
\addplot graphics[
    xmin=20, xmax=120,
    ymin=40, ymax=140
]{
    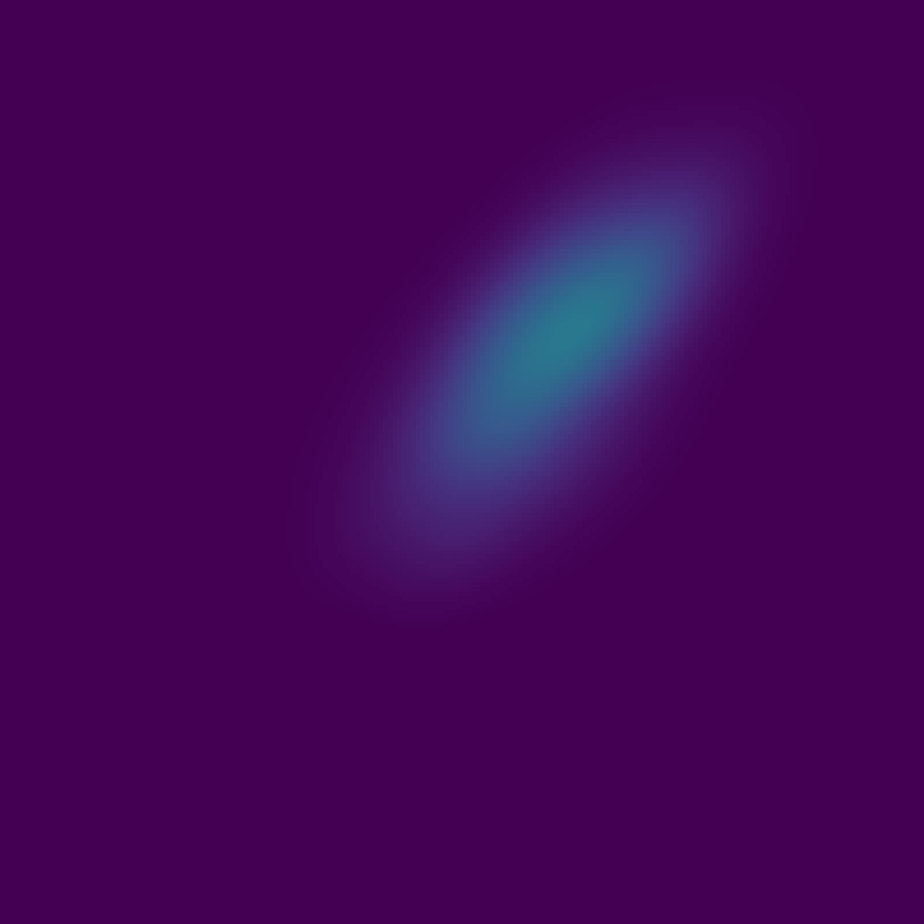
};

\nextgroupplot[densityaxis]
\addplot graphics[
    xmin=20, xmax=120,
    ymin=40, ymax=140
]{
    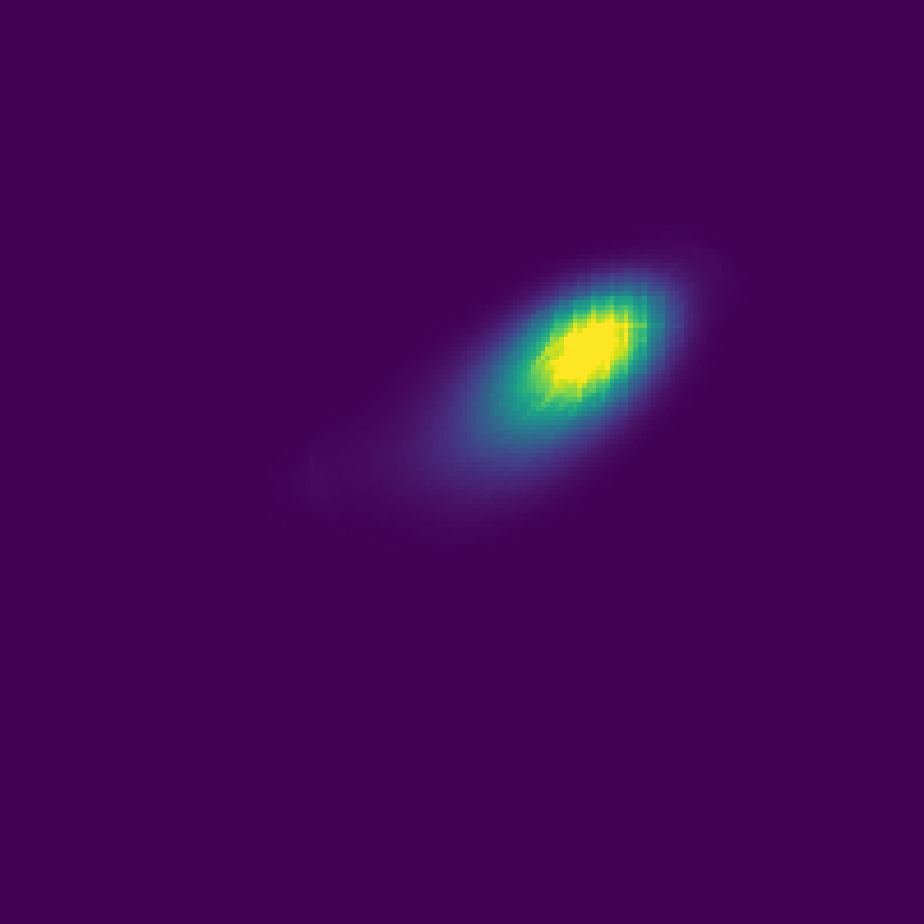
};


\nextgroupplot[emptyaxis]
{}

\nextgroupplot[densityaxis]
\addplot graphics[
    xmin=20, xmax=120,
    ymin=40, ymax=140
]{
    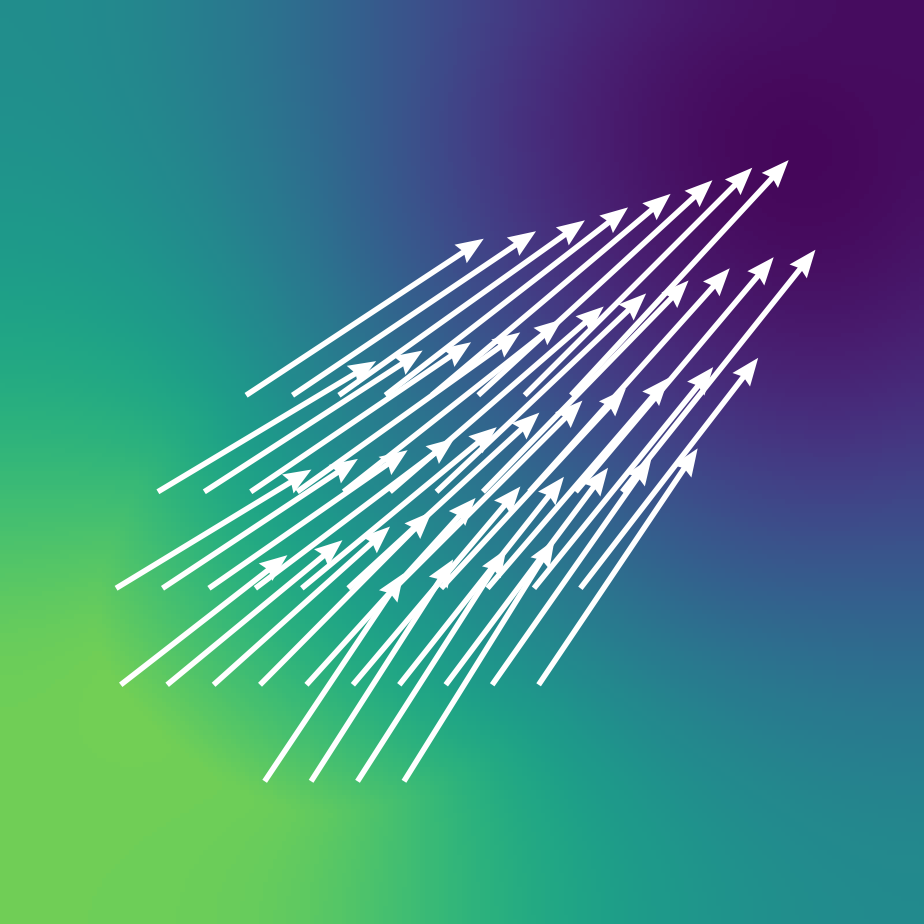
};

\nextgroupplot[densityaxis]
\addplot graphics[
    xmin=20, xmax=120,
    ymin=40, ymax=140
]{
    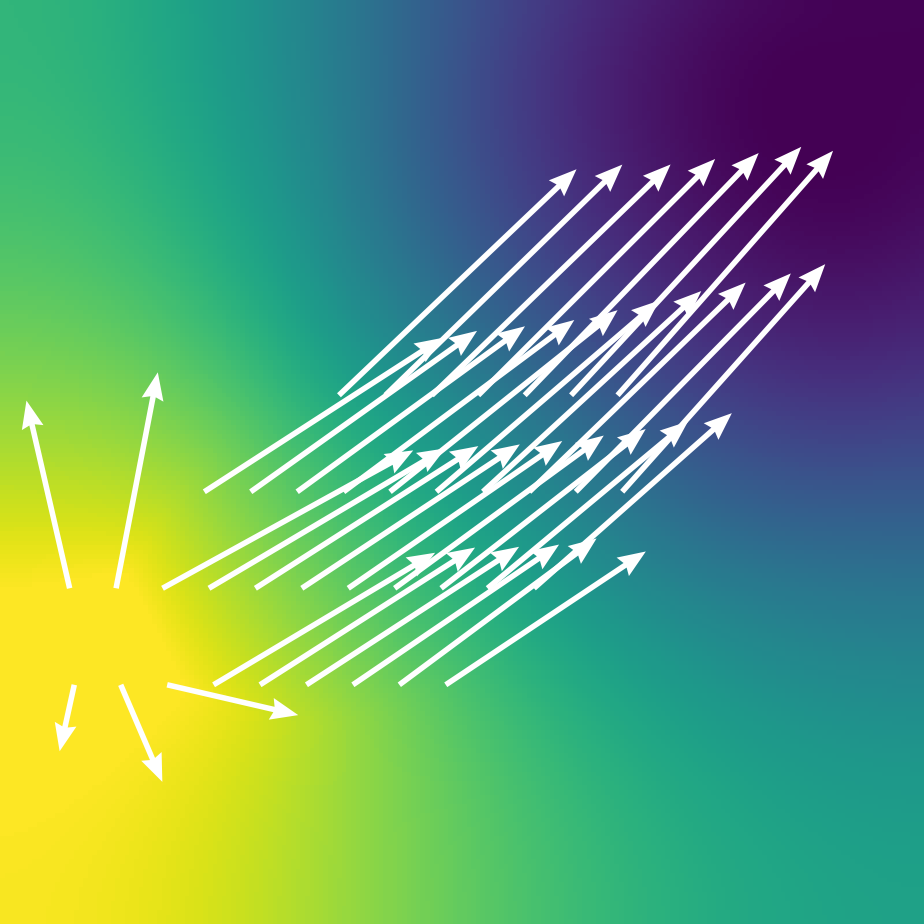
};

\nextgroupplot[densityaxis]
\addplot graphics[
    xmin=20, xmax=120,
    ymin=40, ymax=140
]{
    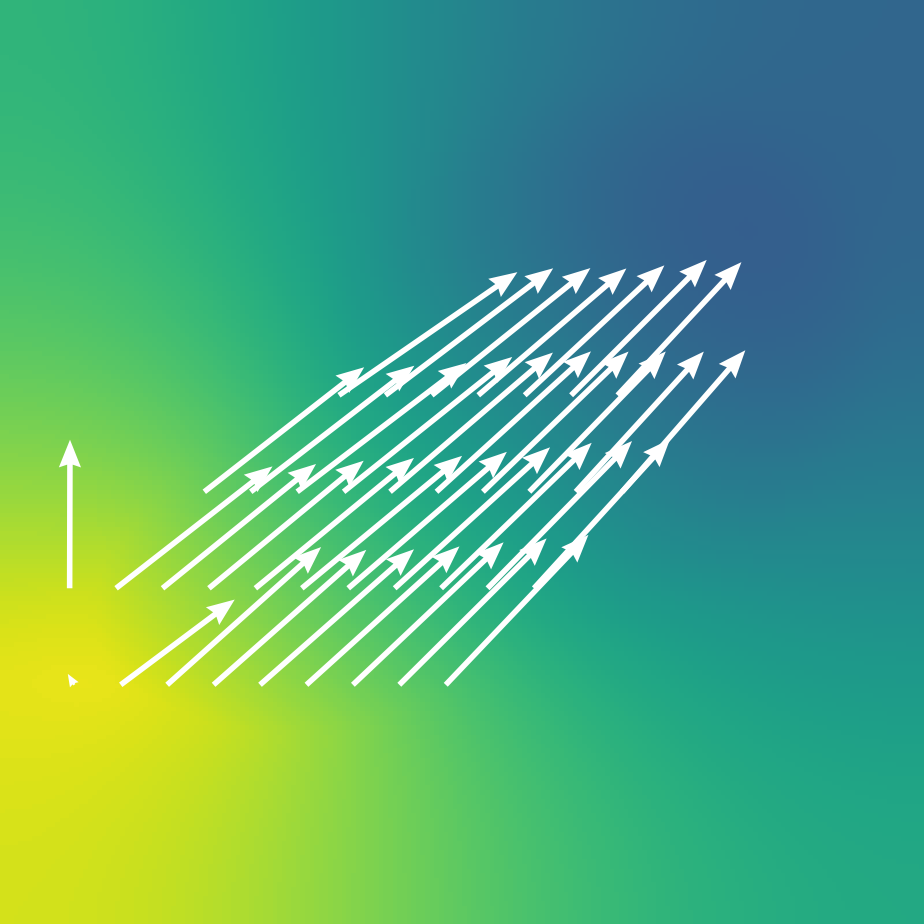
};

\end{groupplot}



\node[anchor=north west, font=\tiny]
at ([xshift=2pt,yshift=-2pt]G c2r1.north west)
{\textcolor{white}{Barycenter in 2000}};

\node[anchor=north west, font=\tiny]
at ([xshift=2pt,yshift=-2pt]G c3r1.north west)
{\textcolor{white}{Barycenter in 2020}};

\node[anchor=north west, font=\tiny]
at ([xshift=2pt,yshift=-2pt]G c4r1.north west)
{\textcolor{white}{Observed 2000}};


\node[anchor=north west, font=\tiny]
at ([xshift=2pt,yshift=-2pt]G c1r2.north west)
{\textcolor{white}{Observed 2020}};

\node[anchor=north west, font=\tiny]
at ([xshift=2pt,yshift=-2pt]G c2r2.north west)
{\textcolor{white}{2020 predicted by GNW}};

\node[anchor=north west, font=\tiny]
at ([xshift=2pt,yshift=-2pt]G c3r2.north west)
{\textcolor{white}{2020 predicted by GOT}};

\node[anchor=north west, font=\tiny]
at ([xshift=2pt,yshift=-2pt]G c4r2.north west)
{\textcolor{white}{2020 predicted by KR}};



\node[anchor=center]
at (G c1r1.center)
{
    \includegraphics[
        height=0.2\textwidth
    ]{
        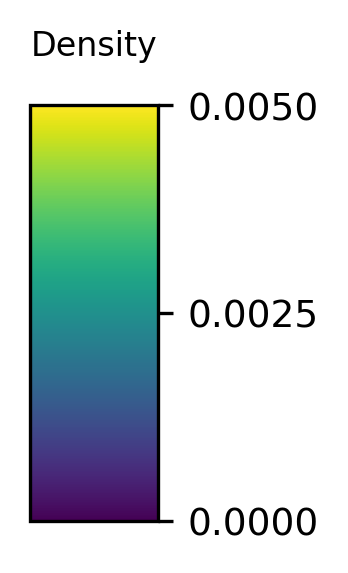
    }
};

\node[anchor=center]
at (G c1r3.center)
{
    \includegraphics[
        height=0.2\textwidth
    ]{
        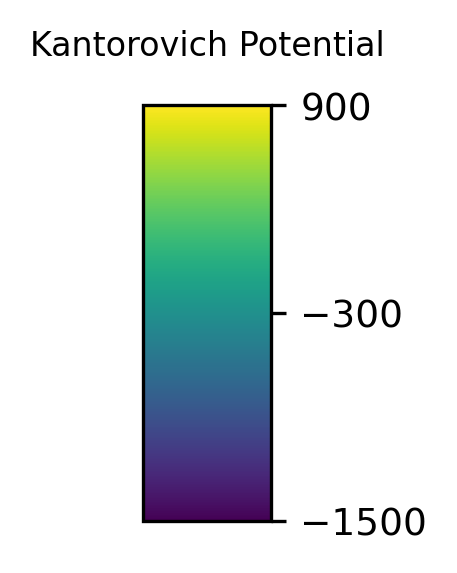
    }
};


\node[anchor=south, font=\small]
at ([yshift=2mm]current bounding box.north)
{Phoenix International Airport, AZ};
\end{tikzpicture}
\caption{
Prediction illustration using Phoenix Airport as the test data. Row 1: Wasserstein barycenters of the temperature distributions in 2000 (col. 2) and 2020 (col. 3), and the observed temperature density for Phoenix Airport in 2000 (col. 4). Row 2: Observed temperature density for Phoenix Airport in 2020 (col. 1) and the corresponding predictions obtained by GNW (col. 2), GOT (col. 3), and KR (col. 4). Row 3: Kantorovich potentials (heat maps) and displacement vectors (white arrows) associated with the observed predictor density (col. 2), observed response density (col. 3), and KR-predicted density (col. 4) for Phoenix Airport.}
\label{fig:pred2D}
\end{figure}

To illustrate the interpretability of the functional parameter, we first fit the model to the entire dataset. The estimated functional parameter $\widehat{f}'$ is plotted in Figure~\ref{2D:sim}. We observe that $\widehat{f}'<1$ over the entire domain. This indicates that the displacement vectors from the Wasserstein barycenter to each distribution have been contracted over the 20-year period, suggesting that the gap between summer temperature distributions across U.S. has decreased. Moreover, the variation in $\widehat f'$ indicates that the contraction is non-uniform across different level sets of the predictor Kantorovich potential, with larger potential values corresponding to stronger contraction. As seen in the bottom row of Figure \ref{fig:pred2D}, larger potential values are associated with lower-temperature regions of the distributions. This pattern suggests that colder regions have experienced stronger warming than hotter regions, thereby reducing differences among summer temperature distributions across the U.S. For further illustration, we used the distribution for Phoenix Sky Harbor International Airport, AZ, as the test data and the distributions for the remaining 49 airports as the training data. The resulting predictions are shown in Figure~\ref{fig:pred2D}. We observe that the location of the probability density predicted by our method is closest to that of the observed response density compared to geodesic optimal transport regression(GOT) [\cite{zhu2025geodesic}] and generalized Nadaraya-Watson regression (GNW) [\cite{gnw}]. The response displacement vector field predicted by our model matches both the direction and the magnitude of the observed response displacement vector.

To evaluate the predictive performance of the competing methods, we conducted leave-one-out cross-validation over the entire dataset. Specifically, each of the 50 distribution pairs was held out in turn as the test set, while the remaining 49 pairs were used to fit the regression models and predict the held-out distribution. Prediction error for each held-out observation was measured by the squared Wasserstein distance between the observed and predicted distributions. We then computed the average and standard deviation of the prediction errors across the 50 leave-one-out iterations. KR achieves the smallest average prediction error of $12.3235$, compared with $12.4718$ for GOT and $24.8800$ for GNW. KR also exhibits lower variability, with a standard deviation of $8.6985$, compared with $17.8486$ for GOT and $46.9810$ for GNW. These results indicate that KR provides both more accurate and more stable predictions than the competing methods.

\printbibliography

\end{document}